\documentclass[preprint,review, 12pt]{elsarticle}

\usepackage{etoolbox}
\newtoggle{paperlayout}
\togglefalse{paperlayout} 

\iftoggle{paperlayout}{

\usepackage{amsmath,amsfonts,amssymb,amsthm}
\usepackage{eurosym}
\usepackage[para]{threeparttable} 
\usepackage{url}
\usepackage{tabularx}
\usepackage{textcomp} 								

\usepackage{color}

}{

\usepackage{fullpage}
\usepackage{times}
\usepackage{epsf}
\usepackage{epsfig}
\usepackage{subfigure}
\usepackage{latexsym}
\usepackage{graphicx}
\usepackage{hyperref}
\usepackage{float}
\usepackage{color}
\usepackage{wrapfig}
\usepackage{topcapt}
\usepackage{multirow}
\usepackage{tabularx}
\usepackage{amsmath, amsfonts, amssymb,amsthm}
\usepackage[hmargin=1in,vmargin=2.5cm]{geometry}
\usepackage{url}
\usepackage{multirow}
\usepackage{textcomp} 								
\usepackage{eurosym}
\usepackage[para]{threeparttable} 

\widowpenalty=1000
\clubpenalty=1000

\input{epsf}

\newcommand{\ba}{\begin{align*}}
\newcommand{\ea}{\end{align*}}

\def\be{\begin{equation}}
\def\ee{\end{equation}}

\definecolor{orange}{rgb}{1,0.5,0}

\setlength{\abovecaptionskip}{-2pt} \setlength{\subfigcapskip}{-2pt}
\setlength{\subfigbottomskip}{6pt} \setlength{\subfigtopskip}{0pt}

}

\newcommand{\unit}[1]{\ensuremath{\mathrm{#1}}}                  

\journal{Acta Astronautica}

\begin{document}

\begin{frontmatter}

\title{On Small Satellites for Oceanography: A Survey}

\author[FCUP]{Andr\'{e} G. C. Guerra\corref{cor1}}
\ead{aguerra@fc.up.pt}

\author[FCUP]{Frederico Francisco} 
\ead{frederico.francisco@me.com}

\author[FEUP]{Jaime Villate}
\ead{villate@fe.up.pt}

\author[VIGO]{Fernando Aguado Agelet} 
\ead{faguado@tsc.uvigo.es}

\author[FCUP]{Orfeu Bertolami}
\ead{orfeu.bertolami@fc.up.pt}

\author[FEUP]{Kanna Rajan} 
\ead{kanna.rajan@fe.up.pt}

\cortext[cor1]{Corresponding author}

\address[FCUP]{Departamento de F\'{i}sica e Astronomia, Centro de F\'{i}sica do Porto, Faculdade de Ci\^{e}ncias, Universidade do Porto, Portugal} %

\address[FEUP]{Faculdade de Engenharia da Universidade do Porto, Portugal} 

\address[VIGO]{Escola de Enxe\~{n}ar\'{i}a de Telecomunicaci\'{o}n, Universidade de Vigo, Espa\~{n}a}

\begin{abstract}

  The recent explosive growth of small satellite operations driven
  primarily from an academic or pedagogical need, has demonstrated the
  viability of commercial-off-the-shelf technologies in space. They
  have also leveraged and shown the need for development of compatible
  sensors primarily aimed for Earth observation tasks including
  monitoring terrestrial domains, communications and engineering
  tests. However, one domain that these platforms have not yet made
  substantial inroads into, is in the ocean sciences. Remote sensing
  has long been within the repertoire of tools for oceanographers to
  study dynamic large scale physical phenomena, such as gyres and
  fronts, bio-geochemical process transport, primary productivity and
  process studies in the coastal ocean. We argue that the time has
  come for micro and nano satellites (with mass smaller than $100\,\unit{kg}$ and 2 to 3 year
  development times) designed, built, tested and flown by academic
  departments, for coordinated observations with robotic assets
  in-situ. We do so primarily by surveying SmallSat missions oriented
  towards ocean observations in the recent past, and in doing so, we
  update the current knowledge about what is feasible in the rapidly
  evolving field of platforms and sensors for this domain. We conclude
  by proposing a set of candidate ocean observing missions with an
  emphasis on radar-based observations, with a focus on Synthetic
  Aperture Radar.

\end{abstract}

\begin{keyword}
Small Satellites, Sensors, Ocean Observation
\end{keyword}

\end{frontmatter}


\section{Introduction}
\label{sec:Introduction}

\begin{figure*}[!hbt]
  \centering
  \includegraphics[width=0.75\textwidth]{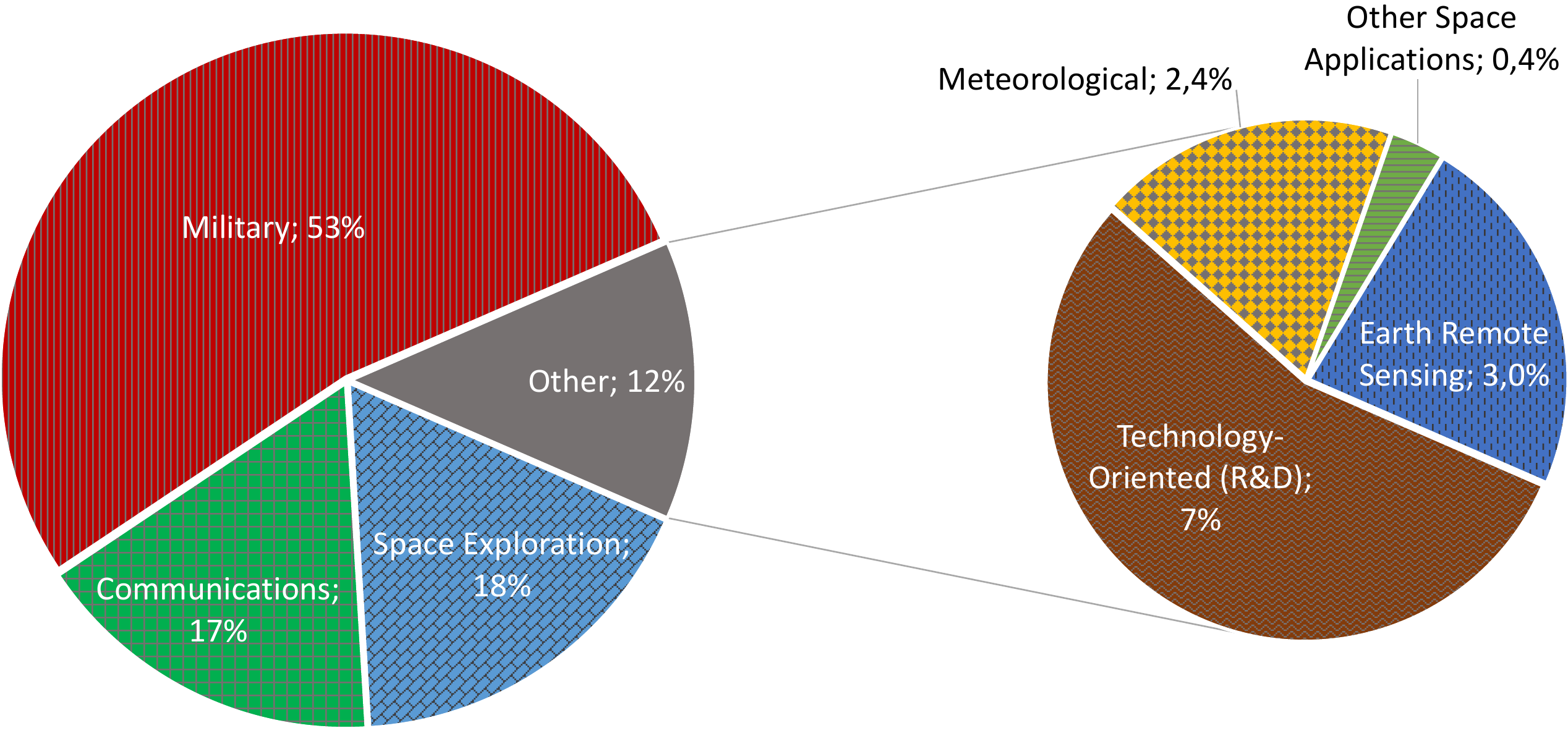}
  \caption{Distribution of the 7472 spacecraft, launched from 1957 to
    2013, according to their mission objectives (\emph{from~\cite{SpacecraftsEncyclopedia}})}
  \label{fig:satdistribution}
\end{figure*}

Starting with \textbf{Sputnik}'s launch in 1957, more than 7000 spacecraft have
been launched, most for communication or military purposes.
Nevertheless, the scientific potential of satellites was perceived
early on, and even though \textbf{Sputnik} did not have any instruments, the
radio beacon it had was used to determine electron density on the
ionosphere~\cite{EONationalResearchCouncilU.S.2008,Kuznetsov2015}. A
small percentage of the satellites were, and still are, dedicated to
research (see Fig.~\ref{fig:satdistribution}). In particular,
we focus on Earth observation and remote sensing satellites, as
they have changed the way we perceive and understand our planet. This
transformation started with the first dedicated weather satellite,
\textbf{TIROS} (Television Infrared Observing Satellite) 1, launched 3 years
after \textbf{Sputnik 1}~\cite{EONationalResearchCouncilU.S.2008}.

Although the first satellites had mass smaller than $200\,\unit{kg}$,
consistent demand on performance led to a natural growth in spacecraft
mass, with direct consequences to their complexity, design, test,
launch, operation and cost. This reached a peak of
$7.9\,\unit{tonnes}$ with ESA's \textbf{EnviSat} mission in
2002~\cite{Selva2012}. With launch costs to low Earth orbit (LEO)
being on average $21\,\unit{k\text{\euro}/kg}$, and for geostationary
Earth orbits (GEO) $29\,\unit{k\text{\euro}/kg}$, for conventional
satellites, the missions were mainly developed by national
institutions or multi-national partnerships involving substantial
investment~\cite{Fortescue2011}. However, several engineering
problems arose from having different instruments (with different
features and requirements) within the confines of a single
spacecraft. Consequently, this rise in mass has stopped and spacecraft
of about $1\,\unit{tonne}$, with fewer instruments have been preferred by
the European Space Agency (ESA) and the American National Aeronautics and Space
Administration (NASA) in the last few years.

The cost of a spacecraft is not only linked to the launch, but
also to its development time, so as to account for mission complexity,
production and operation during its life-span~\cite{Larson1999}.
Moreover, their design, development and subsequent operation require
a substantial infrastructure to provide the end user with the desired data.
Furthermore, project, planning and execution demands years of
investment prior to a successful launch.

The revolution of very-large-scale integration, in 1970, opened the
possibility of integrating sophisticated functions into small volumes,
with low mass and power, which pave the way for the modern small
satellite~\cite{Fortescue2011}. This concept was initially
demonstrated in 1961 with the Orbiting Satellite Carrying Amateur
Radio (\textbf{OSCAR}) 1, and kept growing in sophistication until \textbf{OSCAR-8},
at the end of the 1970s (although still without an on-board
computer). In 1981, the launch of the \textbf{UoSAT-OSCAR-9} (\textbf{UoSAT-1}), of the
University of Surrey, changed this, as it was the first small
satellite with in-orbit re-programmable computers.

In this context, the recent evolution of small satellite (or
SmallSats) designs have proved to be promising for operational remote
sensing. Typically these platforms have come to be classified as
small, micro, and nano satellites. According to the International
Academy of Astronautics (IAA) a satellite is considered small if its
mass is smaller than $1000\,\unit{kg}$~\cite{Sandau2010}. Small
satellites comprise five sub-classes: mini satellites have mass
between $100$ and $1000\,\unit{kg}$; micro satellites, whose mass range
from $10$ to $100\,\unit{kg}$; nano satellites, with mass smaller than
$10\,\unit{kg}$; and satellites with mass smaller than $1\,\unit{kg}$
are referred to as pico satellites~\cite{Sandau2007}. In this work, we
will refer to SmallSats as encompassing all of the above categories.

For each class of mass, there is an expected value for the total cost
(which accounts for the satellite cost, usually 70\% of the total
cost, launch cost, about 20\%, and orbital operations,
10\%~\cite{Fortescue2011}) and developing time, as shown in
Table~\ref{tab:SmallSat_Classification}. Typically, these platforms
have been demonstrated in low to medium Earth orbits, and are launched
as secondary payloads from launch vehicles. Due to this piggyback
launch, sometimes is not the mass of the spacecraft that determines
the launch cost, but the integration with the
launcher~\cite{Swartwout2013b}.

\begin{table}[!htbp]
	\centering
	\caption{Spacecraft classification associated with mass~\cite{Sandau2007}, cost~\cite{Fortescue2011}, and development time~\cite{Briess2011}}
	\label{tab:SmallSat_Classification}
	\iftoggle{paperlayout}{
	\begin{tabular}{cccm{1.8cm}}
	}{
	\begin{tabular}{cccc}
	}
		\noalign{\smallskip}\hline\noalign{\smallskip}
		Satellite Class 	& Mass [kg] 	& Cost [M€]	& Development Time [years]  \\
		\noalign{\smallskip}\hline\noalign{\smallskip}
		Conventional 			& $> 1000$ 		& $> 100$ 	& $> 6$ \\
		Mini  						& 100 - 1000 	& 7 - 100 	& 5 - 6 \\
		Micro 						& 10 - 100 		& 1 - 7 		& 2 - 4 \\
		Nano  						& 1 - 10 			& 0.1 - 1 	& 2 - 3 \\
		Pico  						& $< 1$  			& $< 0.1$ 	& 1 - 2 \\
		\noalign{\smallskip}\hline\noalign{\smallskip}
	\end{tabular}
\end{table}

\subsection{Advantages of SmallSats}

While lower costs are one relevant consideration, SmallSats have other
inherent advantages. With developing times of less than 6 years
(substantially less than larger platforms), SmallSats have more
frequent mission opportunities, and thus, faster scientific and data
return. Consequently, a larger number of missions can be designed,
with a greater diversity of potential users~\cite{Sandau2010}. In
particular, lower costs makes the mission inherently flexible, and less
susceptible to be affected by a single failure~\cite{Shiroma2011}.
Furthermore, newer technologies, can iteratively be applied. This is
particularly pertinent for the adoption of commercial-off-the-shelf
(COTS) microelectronic technologies (i.e. without being space
qualified), and Micro-Electro-Mechanical Systems (MEMS) technology,
that had a powerful impact on the power and sophistication of
SmallSats~\cite{Fortescue2011}. Another important feature of SmallSats
is their capability to link \textit{in situ} experiments with manned ground
stations. As they usually have lower orbits, the power required for
communication is less demanding and there is no need for high gain
antennas, on neither the vehicles nor the ground control
station. Finally, small satellites open the possibility of more
involvement of local and small industries~\cite{Sandau2010}. These
factors combined, make such platforms complementary to traditional
satellites.

There are however some disadvantages associated with small satellites.
The two most conspicuous are the small space and modest power
available for the payload. Both of these have a particular impact on
the instrumentation the spacecraft can carry and the tasks it can
perform. As most of small satellites are launched as secondary
payloads, taking advantage of the
excess launch capacity of the launcher, they, most often, do not have
any control over launch schedule and target orbit~\cite{Swartwout2015}.

Lower cost, greater flexibility, reduced mission complexity and
associated managing costs, make small satellites a particularly
interesting tool for the pedagogical purposes, with a number of
platforms often designed, tested and operated by students (taking the
operations of the spacecraft to the university or even department
level)~\cite{Fortescue2011}. Actually, small satellites projects have
had a substantial educational impact even at the undergraduate student
level. A number of universities over the world, have launched
satellite programmes, and the first launch took place in 1981 (the
already mentioned \textbf{UoSAT-1})~\cite{Swartwout2013}. Rather
unfortunately, these small satellite missions were so different in
kind, in terms of mass, size, power and other features,
that by the turn of the century a series of failures brought
the student satellite missions, especially in the United States, nearly
to a halt. This led to the introduction of a standardisation effort,
that took shape via the CubeSat.

In 1999 Stanford University and the California Polytechnic State
University developed the CubeSat standard, and with it the Poly
Picosatellite Orbital Deployer (POD) for deployment of
CubeSats~\cite{TheCubeSatProgram2015}. The CubeSat (1U) corresponds
to a cube of $10\,\unit{cm}$ side (with a height of $11.35\,\unit{cm}$)
and mass up to $1.33\,\unit{kg}$, even though other sizes are admitted
as a standard (see Table~\ref{tab:CubeSat}). The hardware cost of a
CubeSat can vary between \euro50,000--200,000.

\begin{table}[!htb]
	\centering
	\caption{Standard classes for CubeSats~\cite{TheCubeSatProgram2015}}
	\label{tab:CubeSat}
	\begin{tabular}{ccc}
		\noalign{\smallskip}\hline\noalign{\smallskip}
		Class & Size [cm] 												& Mass [kg] \\
		\noalign{\smallskip}\hline\noalign{\smallskip}
		1U    & 10$\,\times\,$10$\,\times\,$11.35 & $< 1.33$ \\
		1.5U  & 10$\,\times\,$10$\,\times\,$17 		& $< 2$ \\
		2U    & 10$\,\times\,$10$\,\times\,$22.7 	& $< 2.66$ \\
		3U    & 10$\,\times\,$10$\,\times\,$34 		& $< 4$ \\
		3U+   & 10$\,\times\,$10$\,\times\,$37.6 	& $< 4$ \\
		\noalign{\smallskip}\hline\noalign{\smallskip}
	\end{tabular}
\end{table} 

\subsection{Ocean observation}

Spacecraft have become an indispensable tool for Earth Observation 
given their capability to monitor on regional or global scales, and
with high spatial and temporal resolution over long periods of
time~\cite{Garcia-Soto2012}. Several missions have covered and widely
influenced the Earth sciences, from meteorology and oceanography, to
geology and biology~\cite{Kramer1996}. SmallSats (in particular micro
and nano satellites) however, have had little to no impact thus far on
oceanography. We intend to make a case for such a focus.

About $70\%$ of Earth’s surface is covered by the oceans, which are a
fundamental component of the Earth's ecosystem. Variations on the
oceans properties range from: temperature to salinity~\cite{Robinson2010};
the formation or dissolution of episodic
phenomenon such as blooms, fronts or anoxic zones; and anthropogenic
events, such as human induced chemical plumes from oil rigs, ships or
agricultural runoff. Other anthropogenic changes, including increased
pressure on fishing, extended ship traffic due to expanded global
trade, drug and human trafficking and the recent uptick in maritime
disputed boundaries, further call for increased surveillance of the
oceans. One should also point out that the above causes have a
significant impact on the coastal ecosystem, inhabited by about $44\%$
of the human population. Traditional methods for monitoring and
observation have involved static assets such as moorings with sensors
fixed to the ocean floor, Lagrangian drifters or subsampling by
manned ships or boats. Understanding the change of ocean features has
generated the need for \emph{synoptic} measurements with observations
over larger spatial and temporal scales, critical to deal with
subsampled point measurements. More recently, robotic platforms such
as autonomous underwater vehicles (AUVs), including slower moving
gliders, have extended the reach of such traditional methods. As a
consequence, scientists are now able to characterise a wider swath of
a survey area in less time and more cost effectively. While an
advance, they are not considered to be spatio-temporally synoptic,
since they do not match the mesoscale ($> 50\,\unit{km^2}$) observation
capability, necessary to digest the evolving bio-geochemistry or for
process studies, especially in the coastal ocean.

On the other hand, synthetic ocean models to provide means for
prediction also come with their own limitations. Their \emph{skill
level}, in particular, is often poor given the complexity of mixing
in near shore waters. Therefore, observation prediction continues to be
a challenge hampering a better understanding of the global
oceans. Such a need is demonstrably important in situations like
oil spill response, such as the Macando event in the Gulf of Mexico.

With recent advances in unmanned aerial vehicles (UAVs), payload
sensors, and their cost-effectiveness in field operations, these
platforms offer a tantalising hope of further extending the reach of
oceanographers. Nevertheless, ship-based or robotic-based observation
and surveillance methods have yet to provide that level of scale,
maturity or robustness (or all of the above) appropriate for providing
rapid maritime domain and situation awareness.

For all these reasons, satellite remote sensing continues to be a
valuable tool, providing a range of high-resolution data including
imagery. More recently, radar imagery, from LIDAR (Light Detection and
Ranging) to SAR (Synthetic Aperture Radar), have been a boon for
maritime authorities, as well as policy makers, with their clear
all-weather capability to observe the oceans. The costs and complexity
to provide such capabilities however is a significant shortcoming,
especially in the study of the changing ocean.

\subsection{SmallSats and Ocean Observations}

Our thesis is that micro and nano satellites are a
key element in the (near) future needs of oceanography. Besides the
cost and more modest operational requirements, increasingly smaller
form-factor sensors are being designed and built for autonomous
robotic (terrestrial, aerial and underwater) platforms, which can be
leveraged for SmallSats. Furthermore, as sensors become increasingly
affordable, we can envision \emph{multiple} SmallSats in a
constellation with identical sensors, pointing Earthwards, to provide
near real-time coverage to any part of the planet, especially the
remote oceans. Launch costs associated with a single SmallSat are
unlikely to be substantially reduced from multiply launched
vehicles. In other words, we envisage that multiple SmallSats carrying
appropriate sensors in the same orbital plane, can and should become
an extension of oceanographic sensing. While such platforms cannot
provide \textit{in situ} sampling capability, their synoptic observations can
be used to intelligently provide such a capability to ensure robotic
elements are ``at the right place and right time'', to obtain data
especially of episodic phenomenon. 

Thus, combining the robotic capabilities of autonomous underwater
vehicles (AUVs), autonomous surface vehicles (ASVs), and unmanned
aerial vehicles (UAVs), with dedicated small satellites will
considerably boost the study of the oceans, provide synoptic views
potentially close to real-time, and yield an unprecedented view of
our evolving ecosystem on Earth, characterised by the large mass of the
ocean. SmallSats, AUVs, ASVs and UAVs are then elements in a strategy
to provide \emph{coordinated observation} data. Another key objective
of our work, is to show that such a capability, along with novel
methods of multi-vehicle control, can provide a new observational
capacity, that combines both hardware and software so to make
data more accessible to the oceanographer, at whatever scales one
wishes to select. These can be from small temporal scales of observing evolving
harmful algal blooms, to large spatial scales of movement of coherent
Lagrangian structures~\cite{haller2015}. Together with new methods in
data science and analytics, trends can be examined in more detail,
with point measurements giving way to near continuous observations
of the ocean surface and potentially upper water-column. And to be
able to do so in a cost-effective manner.

Our objective in this survey paper is to provide a timely and
comprehensive view, with a wide perspective, of what SmallSats
capabilities are currently available, and in many cases previously
proposed. And in surveying the field rigorously, we attempt to provide
a perspective of the use of such technologies now, and in the near
future, for oceanographic measurements. Our intended audience is
scientists and engineers typically in the ocean sciences, and students
in engineering with interests in making ocean measurements and ocean
engineering, especially marine robotics.

This paper is organised as a survey of SmallSat missions for
oceanography cited in the literature, coupled with current methods in
robotic oceanographic observations, and SmallSat characteristics, to
make a strong case for their focus on this domain. The organisation
of this paper is as follows: in Section~\ref{sec:robots}, we motivate
the use of SmallSats with \textit{in situ} robotic platforms; in
Section~\ref{sec:Survey}, we survey prominent examples of micro and nano satellites
and, in particular, those designed for oceanographic studies or
monitoring; in Section~\ref{sec:Components_Features}, we discuss the
main building blocks of SmallSats; and in
Section~\ref{sec:Sensors_Oceanography}, we present a common set of
sensors useful for oceanography. Finally, in
Section~\ref{sec:discuss}\ we conclude with a discussion and our
conclusions and perspectives.


\section{Motivation: Coordinated Observations with autonomous platforms}
\label{sec:robots}

Observing the ocean synoptically in space and time is an increasingly
important goal facing oceanographers. The changing climate has become
a major societal problem to tackle, and with an emphasis on the ocean
as the primary sink for greenhouse gases, ocean science and the study
of the changing climate has become critical to understanding our
planet. However, much has been made of the lack of data and the need
to analyse oceanographic phenomena at large scales, not approachable
with current observation tools and methodologies, that often rely on
traditional ship-based methods. To study phenomena and processes in
the ocean which can spatially and temporally range from minutes to
months, ship-based methods are neither cost-effective nor sustainable
in the long run. Further such measurements do not provide a continuous
scale of change in either space or time necessary to understand
natural variability.

\begin{figure}[!ht]
  \centering
	\iftoggle{paperlayout}{
	\includegraphics[width=1\columnwidth]{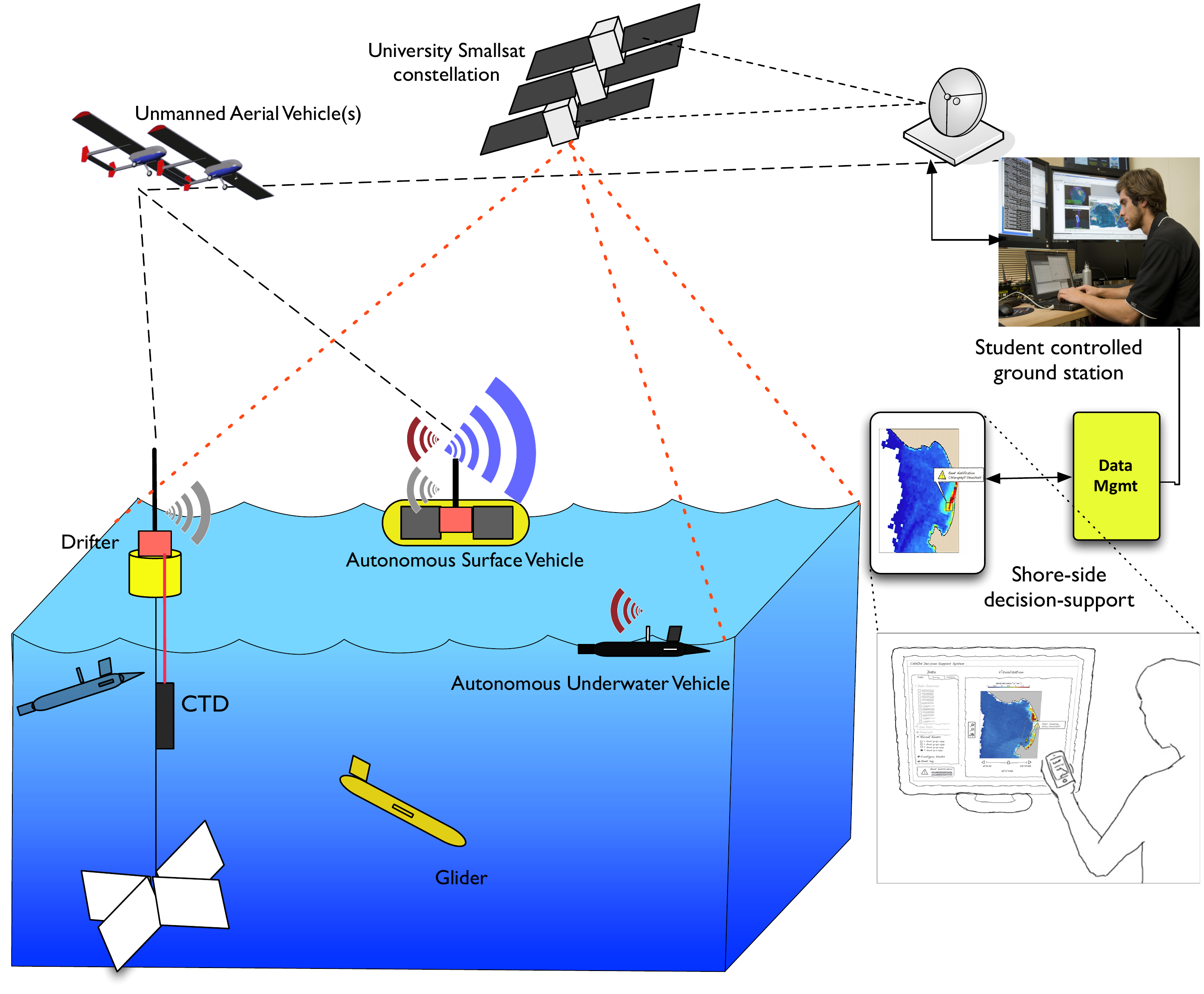}
	}{
	\includegraphics[scale=0.5]{fig/networked.pdf} 
	}
  \caption{Networked robotic platforms for synoptic oceanographic observations.}
  \label{fig:ssats}
\end{figure}

Recent advances in robotic vehicles have made a dent in more
sustainable ocean observation, with the use of autonomous and
semiautonomous platforms to observe at such varying spatio-temporal
scales. While this is a good start, we are at the infancy of
systematic observations using the current generation of robotic
hardware. The principal challenge is to observe a water column not
just with point-based observations, but across the mediums of air and
water, and the air/water interface, and doing so continuously. Such
observations need not only to be synoptic, but also to be coordinated
across space and time to observe the same patch of the ocean at the
same time. Its enabling requires coordination and control of a range
of robots with appropriate sensors, across the space, aerial, surface
and underwater domains (Fig.~\ref{fig:ssats}).

This necessitates the use of multi-platform systems to observe a patch
of the ocean in the meso-scale ($> 50\,\unit{km^2}$), and to follow
targeted phenomena of interest such as blooms, plumes, anoxic zones,
fronts over a period of days, weeks or longer. Recent
experiments~\cite{ryan10,Ryan2013,pinto13,pinto14,das15} have led to
the conclusion that multiple vehicles, operating in different
operational domains, are critical for such observational
needs. Detection of the targeted features however require observation
over larger scales, with the consequent need for remote sensing data
to drive robotic assets \textit{in situ}.

Current generation of Earth (and ocean) observing spacecraft provide a
rich trove of remote sensing data, and have been credited with
understanding the anthropogenic impact on our environment
\cite{ipcc14}. These systems have been complex, expensive and years in
the making for design, build, test and launch operations, and
typically undertaken by governmental agencies or multinational
institutions.

The advent of the smartphone has resulted in a revolution in sensor
technology with a corresponding surge in applications for
robotics. This surge has also seeped into satellite technology, with
SmallSats being produced as student projects, and operated in a far
more affordable way. SmallSats provide just such a novel approach to
augmenting ocean observation methods. Their coordination with Earth
bound robotic platforms, with all-weather radar imagery, will allow
for localising features or targets (in security scenarios) for the
ensemble of vehicles in the open ocean. Such experiments have already
commenced~\cite{pinto13,pinto14}, albeit without the space
component\footnote{See
  \url{http://rep13.lsts.pt/},\url{http://rep15.lsts.pt/} and
  \url{http://sunfish.lsts.pt/}}. A range of applications, starting
with upper water-column observations targeting Harmful Algal Blooms,
plumes, anoxic zones and frontal zones, will provide a rich trove of
experience and equally importantly, scientific data. Developments
along these lines, will push the state of the art, and practice, in
both engineering methods in marine robotics as well as in ocean
science. Moreover, the technology is dual-use and has clear
applications to maritime security and defence.

Equally important, SmallSats are increasingly viable as student-led
projects. Not only do they provide a vehicle for student involvement,
for a successful operation, such projects provide students with a
strong background in systems engineering and collaborative work in
teams across various disciplines, including electronics, material
science, physics and control systems to name a few.

We call for an increased focus in SmallSats, both as a technological
push, and for obtaining new and sustained methods of observation and
data collection, as also as pedagogical tools for students in
inter-disciplinary science and engineering. Student-led missions, with
help from various departments, will be required to design materials,
sensors and software to be embedded within these space-borne
assets. Such platforms can then be part of a robotic ensemble, which
can be controlled for purposes of making sustained coordinated
measurements in offshore environments.


\section{Survey of Ocean Observing SmallSats}
\label{sec:Survey}

We focus now on a survey of SmallSats in the context of observing the
oceans. In this and following sections, we categorise SmallSats
in terms of micro, nano or CubeSats. The survey covers those
in design stages, successfully operating, as well as those which were
unsuccessful in meeting their operational requirements. This survey is
up to date until around Fall 2015.

Some of the fields where SmallSats (especially below $100\,\unit{kg}$)
have proved useful are ocean imaging, data storage and relay, and
traffic monitoring (e.g. through the Automatic Identification System,
AIS). While not all the surveyed examples involve oceanographic
variables per se, they have relevant bearing on oceanography. We also
consider missions where a
constellation of SmallSats is employed.

The SmallSat's described here were referenced through available online
databases (the Earth Observation
Portal\footnote{\url{https://directory.eoportal.org/web/eoportal/satellite-missions}},
the Union of Concerned Scientist satellite
database\footnote{\url{http://www.ucsusa.org/nuclear-weapons/space-weapons/satellite-database.html}},
and the Nanosatellite
Database\footnote{\url{http://www.nanosats.eu/}}), and via a wider
survey of the literature as well as the web. Whenever possible, references of the different
missions were used, although some information comes from the
databases. Table~\ref{tab:SmallSat_Survey} summarises the spacecraft
discussed here, with mass, power, mission and payload (related to
oceanography), and launch date (past and future).

\iftoggle{paperlayout}{
\begin{table*}[!htb]
  \centering
	\caption{Summary of Ocean Observing SmallSats (Operational if not stated otherwise)}
	\label{tab:SmallSat_Survey}
	\footnotesize
		\begin{threeparttable}
    \begin{tabular}{m{2cm}cccm{5.5cm}m{3cm}c}
			\noalign{\smallskip}\hline\noalign{\smallskip}
			Satellite 									& Mass [kg]	& Power [W] & Size [cm]\tnote{a}							& Mission\tnote{b}																		& Payload\tnote{b}							& Launch \\
			\noalign{\smallskip}\hline\noalign{\smallskip}
			ZACube-2\tnote{$\ddagger$}	& 4         & --    		&	10$\,\times\,$10$\,\times\,$34	& Vessel tracking	and ocean colour										& AIS and imager								& -- \\
			RISESat\tnote{$\ddagger$} 	& 60    		& 100   		& 50$\,\times\,$50$\,\times\,$50	& Fisheries and environment studies										& Multi-band camera 						& 2016 \\ 
			M3MSat\tnote{$\ddagger$} 		& 85    		& 80    		& 80$\,\times\,$60$\,\times\,$60	& Augmenting maritime surveillance capabilities				& AIS														& 2016 \\
			AAUSat5 										& 0.88 			& 1.15  		& 10$\,\times\,$10$\,\times\,$11	& Vessel tracking   																	& AIS														& 2015 \\
			LambdaSat\tnote{$\star$} 		& 1.5       & 1.5   		& 10$\,\times\,$10$\,\times\,$11	& Vessel tracking																			& AIS														& 2015 \\
			LAPAN-A2 										& 68    		& 32    		& 50$\,\times\,$47$\,\times\,$36	& AIS payload for the equatorial region								& AIS														& 2015 \\
			AISat 											& 14        & 15    		& 10$\,\times\,$10$\,\times\,$11	& Helical antenna technology demonstration						&	AIS														& 2014 \\
			QSat-EOS 										& 50    		& 70    		& 50$\,\times\,$50$\,\times\,$50	& Ocean eutrophication																& VIS and NIR camera\tnote{c}		& 2014 \\ 
			AAUSat3\tnote{$\star$} 			& 0.8       & 1.15  		& 10$\,\times\,$10$\,\times\,$11	& Vessel tracking 																		& AIS 													& 2013 \\
			COPPER\tnote{$+$}	 					& 1.3   		& 2.5   		& 10$\,\times\,$10$\,\times\,$11	& Infrared Earth images																& Uncooled Microbolometer Array	& 2013 \\ 
			WNISAT-1 										& 10        & 12.6  		& 27$\,\times\,$27$\,\times\,$27	& Monitoring Arctic Sea state													& VIS and NIR cameras						& 2013 \\ 
			Aeneas 											& 3     		& 2     		&	10$\,\times\,$10$\,\times\,$34	& Track cargo containers 															& Antenna and corresponding electronics	& 2012 \\ 
			VesselSat 2 								& 29    		& --    		& 30$\,\times\,$30$\,\times\,$30	& Collect space-borne AIS message data								& AIS														& 2012 \\
			SDS-4 											& 50    		& 60    		& 50$\,\times\,$50$\,\times\,$45	& Demonstrate space-based AIS													& AIS 													& 2012 \\ 
			exactView 1 								& 98    		& --    		& 63$\,\times\,$63$\,\times\,$60	& Commercial constellation of AIS spacecraft					& AIS														& 2012 \\
			exactView 6/5R/12/11/13 		& 13    		& 15    		& 25$\,\times\,$25$\,\times\,$25	& Commercial constellation of AIS spacecraft					& AIS														& 2011/13/14 \\
			VesselSat 1 								& 29    		& --    		& 30$\,\times\,$30$\,\times\,$30	& Collect space-borne AIS message data								&	AIS														& 2011 \\
			AISSat-1  									& 6.5   		& 0.97  		&	20$\,\times\,$20$\,\times\,$20	& Assess feasibility of situational awareness service	& AIS														& 2010 \\
			NTS/CanX-6\tnote{$\star$}		& 6.5   		& 5.6   		&	20$\,\times\,$20$\,\times\,$20	& Demonstrate AIS detection technology								& AIS														& 2008 \\
			WEOS  											& 50        & 14.5			& 52$\,\times\,$52$\,\times\,$45	& Data relay from cetacean probes											& UHF Antenna 									& 2002 \\ 
			\noalign{\smallskip}\hline\noalign{\smallskip}
    \end{tabular}
		\begin{tablenotes}
			\item [a] Size of the main structure (not accounting for deployed mechanisms, e.g. antennas).\\
			\item [b] Related to oceanography.\\
			\item [c] Visible (VIS) and near infrared (NIR).\\
			\item [$\ddagger$] To be launched.\\
			\item [$\star$] Concluded operations.\\
			\item [$+$] Mission failed.
		\end{tablenotes}
		\end{threeparttable}
\end{table*}
}{
\begin{table}[!htb]
  \centering
	\caption{Summary of Ocean Observing SmallSats (Operational if not stated otherwise)}
	\label{tab:SmallSat_Survey}
	\footnotesize
	\scalebox{0.85}{
		\begin{threeparttable}
    \begin{tabular}{m{2cm}cccm{5.5cm}m{3cm}c}
			\noalign{\smallskip}\hline\noalign{\smallskip}
			Satellite 									& Mass [kg]	& Power [W] & Size [cm]\tnote{a}							& Mission\tnote{b}																		& Payload\tnote{b}							& Launch \\
			\noalign{\smallskip}\hline\noalign{\smallskip}
			ZACube-2\tnote{$\ddagger$}	& 4         & --    		&	10$\,\times\,$10$\,\times\,$34	& Vessel tracking	and ocean colour										& AIS and imager								& -- \\
			RISESat\tnote{$\ddagger$} 	& 60    		& 100   		& 50$\,\times\,$50$\,\times\,$50	& Fisheries and environment studies										& Multi-band camera 						& 2016 \\ 
			M3MSat\tnote{$\ddagger$} 		& 85    		& 80    		& 80$\,\times\,$60$\,\times\,$60	& Augmenting maritime surveillance capabilities				& AIS														& 2016 \\
			AAUSat5 										& 0.88 			& 1.15  		& 10$\,\times\,$10$\,\times\,$11	& Vessel tracking   																	& AIS														& 2015 \\
			LambdaSat\tnote{$\star$} 		& 1.5       & 1.5   		& 10$\,\times\,$10$\,\times\,$11	& Vessel tracking																			& AIS														& 2015 \\
			LAPAN-A2 										& 68    		& 32    		& 50$\,\times\,$47$\,\times\,$36	& AIS payload for the equatorial region								& AIS														& 2015 \\
			AISat 											& 14        & 15    		& 10$\,\times\,$10$\,\times\,$11	& Helical antenna technology demonstration						&	AIS														& 2014 \\
			QSat-EOS 										& 50    		& 70    		& 50$\,\times\,$50$\,\times\,$50	& Ocean eutrophication																& VIS and NIR camera\tnote{c}		& 2014 \\ 
			AAUSat3\tnote{$\star$} 			& 0.8       & 1.15  		& 10$\,\times\,$10$\,\times\,$11	& Vessel tracking 																		& AIS 													& 2013 \\
			COPPER\tnote{$+$}	 					& 1.3   		& 2.5   		& 10$\,\times\,$10$\,\times\,$11	& Infrared Earth images																& Uncooled Microbolometer Array	& 2013 \\ 
			WNISAT-1 										& 10        & 12.6  		& 27$\,\times\,$27$\,\times\,$27	& Monitoring Arctic Sea state													& VIS and NIR cameras						& 2013 \\ 
			Aeneas 											& 3     		& 2     		&	10$\,\times\,$10$\,\times\,$34	& Track cargo containers 															& Antenna and corresponding electronics	& 2012 \\ 
			VesselSat 2 								& 29    		& --    		& 30$\,\times\,$30$\,\times\,$30	& Collect space-borne AIS message data								& AIS														& 2012 \\
			SDS-4 											& 50    		& 60    		& 50$\,\times\,$50$\,\times\,$45	& Demonstrate space-based AIS													& AIS 													& 2012 \\ 
			exactView 1 								& 98    		& --    		& 63$\,\times\,$63$\,\times\,$60	& Commercial constellation of AIS spacecraft					& AIS														& 2012 \\
			exactView 6/5R/12/11/13 		& 13    		& 15    		& 25$\,\times\,$25$\,\times\,$25	& Commercial constellation of AIS spacecraft					& AIS														& 2011/13/14 \\
			VesselSat 1 								& 29    		& --    		& 30$\,\times\,$30$\,\times\,$30	& Collect space-borne AIS message data								&	AIS														& 2011 \\
			AISSat-1  									& 6.5   		& 0.97  		&	20$\,\times\,$20$\,\times\,$20	& Assess feasibility of situational awareness service	& AIS														& 2010 \\
			NTS/CanX-6\tnote{$\star$}		& 6.5   		& 5.6   		&	20$\,\times\,$20$\,\times\,$20	& Demonstrate AIS detection technology								& AIS														& 2008 \\
			WEOS  											& 50        & 14.5			& 52$\,\times\,$52$\,\times\,$45	& Data relay from cetacean probes											& UHF Antenna 									& 2002 \\ 
			\noalign{\smallskip}\hline\noalign{\smallskip}
    \end{tabular}
		\begin{tablenotes}
			\item [a] Size of the main structure (not accounting for deployed mechanisms, e.g. antennas).\\
			\item [b] Related to oceanography.\\
			\item [c] Visible (VIS) and near infrared (NIR).\\
			\item [$\ddagger$] To be launched.\\
			\item [$\star$] Concluded operations.\\
			\item [$+$] Mission failed.
		\end{tablenotes}
		\end{threeparttable}
	}
\end{table}
}

\subsection{Ocean Imaging}
\label{subsec:Ocean_Imaging}

While a number of imaging nano and micro satellites have been flown,
and although some of the SmallSat imagery could be used for
oceanographic studies, only a few are actually focused on ocean
observation. No CubeSats were found that were equipped with ocean
imaging systems~\cite{Selva2012}, although in 2013 \textbf{COPPER} was
about to be the first. Among micro satellites, only two were found
that have already been launched: one for commercial purposes
(\textbf{WINISAT-1}); the other (\textbf{QSat-EOS}) for monitoring
ocean health, using data acquired for other purposes.


\paragraph{\textbf{RISESat}} The Rapid International Scientific
Experiment Satellite (\textbf{RISESat}) is being developed under an
international cooperation led by Tohoku University, within the
Japanese programme FIRST (Funding Program for World-Leading Innovative
R\&D on Science and Technology)~\cite{Kuwahara2012}. With a launch
planned for 2016, the objectives are two fold: technological -- to
demonstrate the platform performance, as it is planned to be a common
bus for future missions; and scientific -- through the integration of scientific
payloads from different countries, mainly focused on the Earth and its
environments (to a total mass of about
$10\,\unit{kg}$)~\cite{RISESat_EOPortal}.

The spacecraft bus, the Advanced Orbital Bus Architecture (AOBA), is
intended to be versatile, cost effective, and with a short development
schedule, to be compatible (and competitive) for future scientific
missions. The size of the bus is expected to be smaller than
50$\,\times\,$50$\,\times\,$50 cm, with a maximum mass equal to
$60\,\unit{kg}$ (but typical less than $55\,\unit{kg}$). The main
structure has a central squared pillar, with one side extended to
connect to the outer panels (increasing the mounting surface and
resulting in a more stable satellite). The outer panels are made of
aluminium isogrid.

Attitude and orbit determination is performed through the use of two
star sensors, a 3-axis fibre optic gyroscope, a 3-axis magnetometer, a GPS receiver, and
coarse and accurate sun sensors (covering 4$\pi$). Control is achieved
with four reaction wheels and 3-axis magnetic torque rods. To achieve
a high reliability, some sensors and actuators are designed with
redundancy.

Power is supplied by gallium arsenide (GaAs) multi-junction cells,
divided in two deployable panels and a body-mounted one, generating
more than $100\,\unit{W}$. The enhanced amount of electric power,
around double the power consumption (expected to be more than
$50\,\unit{W}$), allows for long period and multi instrument
observations. A power control unit (PCU) provides each subsystem
supply lines, although, some more demanding components (in terms of
power) are also directly connected to the PCU. Nickel–metal hydride
(NiMH) type batteries are also installed.

Three different bands are used for communications. The UHF is used for
command uplinks, and a S-band for housekeeping downlink. As large
amounts of scientific data are expected to be generated, a X-band
downlink system was added~\cite{RISESat_EOPortal}.

Furthermore, a de-orbit mechanism will be installed, so to make the
spacecraft re-enter Earth in about 25 years, following the standards
for the maximum re-entry time after the mission
completion~\cite{Klinkrad2004,Taylor2007}. Six micro cameras will
also be installed to monitor the satellite's structure and deployment,
and to view the Earth and where the instruments are pointing.

The payloads include a High Precision Telescope, a Dual-band Optical
Transient Camera, a Ocean Observation Camera (OOC), a
Three-dimensional Telescope, a Space Radiation micro-Tracker, a
micro-Magnetometer, a Very Small Optical Transponder, and a Data
Packet Decoder.

The only focused in ocean observation, the OOC is a multi-band camera, with about
$100\,\unit{m}$ spatial resolution, and a wide field of view (swath width of
approximately $65\,\unit{km}$)~\cite{Kuwahara2012}. Although the
system will work in a continuous acquisition mode, the region of
interest is around Japan and Taiwan. The resulting data will mainly be
used for fisheries and environmental studies. The instrument is
compatible with Space Plug \& Play technology, and has a mass of
about $1\,\unit{kg}$~\cite{RISESat_EOPortal}. Furthermore, it has a
power consumption of less than $3\,\unit{W}$ ($5\,\unit{V}$), and uses
a Watec CCD (charge-coupled device) with 659$\,\times\,$494 pixels
(square pitch size of $7.4\,\unit{\mu m}$), for each of the three
F/1.4 lens.


\paragraph{\textbf{QSat-EOS}} Graduate students of Kyushu University,
in Japan, started developing a spacecraft for space science. However,
the main objective of the mission was shifted to Earth Observation,
in particular to disaster monitoring, due to funding
requirements~\cite{QSatEOS_EOPortal}. Other objectives include
monitoring of Earth's magnetic field, detection of micro debris, and
observation of water vapour in the upper
atmosphere~\cite{Aso2011}. Data acquired also supports other studies,
namely agricultural pest control, red tide (harmful algal bloom)
detection, and ocean eutrophication (pollution due to excess of
nutrients~\cite{Wallace2014}). Launched in 2014, data is currently
being gathered by this SmallSat.


\paragraph{\textbf{COPPER}} The Close Orbiting Propellant Plume and
Elemental Recognition CubeSat was an experimental mission to study the
ability of commercially available compact uncooled microbolometer
detector arrays to take infrared images, besides providing space
situational awareness~\cite{Massey2012}. Additionally, it was intended
to improve models of radiation effects on electronic devices in
space. This SmallSat was part of the Argus programme, a proposed
flight programme of a few CubeSat spacecrafts spanning over many
years, of which \textbf{COPPER} was the pathfinder
mission~\cite{COPPER_EOPortal}. It was launched in November 2013, but
communications were unable to be established.


\paragraph{\textbf{WNISAT-1}} One of the largest weather companies,
Weathernews Inc. of Tokyo, funded the sea monitoring mission
\textbf{WNISAT-1} (Weathernews Inc. Satellite-1), which was launched
in 2013~\cite{AXELSPACECooperation2013}. The objective was to provide
data on the Arctic Sea state, in particular ice coverage, to shipping
customers operating on that area~\cite{Kim2010}. The micro satellite
was designed by a university venture company, AXELSPACE. The aim was
to use a simple architecture and, when possible, COTS devices, with a
compact design, simple operational support, and minimum
redundancies. This SmallSat continues to be operational.

\subsection{Data Relay SmallSats}
\label{subsec:Data_Relay}

There are various missions that do not have payloads designed
specifically for oceanographic studies, but even so they perform or
support work in oceanography, as data relay spacecraft. Some
receive data from \textit{in situ} experiments (e.g. buoys), and transmit the
data received when a ground station is in view. This alleviates the
need for a human operator to monitor an experiment, and to get the
acquired data. A prominent example is the Whale Ecology Observation
Satellite (\textbf{WEOS}) of the Chiba Institute of Technology in
Japan.

Other examples within the realm of data relay SmallSats are the
\textbf{ParkinsonSat}~\cite{ParkinsonSat_EOPortal}, the
\textbf{TUBSAT-N} and \textbf{N1} nano spacecrafts (which were
launched from a submarine)~\cite{TUBSAT_EOPortal}, the twin
\textbf{CONASAT} $8.2\,\unit{kg}$ spacecrafts~\cite{Carrara2014}, and
the \textbf{FedSat} (Federation Satellite)~\cite{Fedsat_EOPortal}.

The submarine launch, the first commercially, was made from a Russian
vessel, using a converted ballistic missile, in 1998. Such
submarine-based launches open up the possibility of sending a
spacecraft to any orbit inclination, without requiring any space
manoeuvres. Clearly, form-factors are critical, since only small
spacecraft can be launched in this manner, due to size and rocket power
constraints.


\paragraph{\textbf{WEOS}} Launched in 2002 to a Sun-synchronous orbit
of about $800\,\unit{km}$, the \textbf{WEOS} spacecraft was designed,
built and operated by students of the Chiba Institute of Technology,
in Japan. The goal was to track signals emitted by probes attached to
whales, while studying their migration routes. The probe transmitted
GPS position, diving depth, and sea temperature using the UHF
band~\cite{Hayashi2004}. Although the spacecraft is still operational,
scientists were unable to attach the probes to whales, and only
tests with ocean buoys were made.

\subsection{Tracking and AIS}
\label{subsec:AIS}

Maritime domain awareness and security is a key need for governments
and policy makers worldwide. This includes protection of critical
maritime infrastructures, enforcement of the freedom of navigation,
deterrence, preventing and countering of unlawful
activities~\cite{EU_MaritimeSecurity2014}. A number of countries are
paying special attention to this problem. One example is Canada, where
1600 ships transect its extended continental shelf per day~\cite{Bedard2007}.
Japan, Norway and Indonesia are other examples of countries
paying attention to maritime traffic, given that their exclusive
maritime economic zones are
substantial~\cite{Eriksen2010a,Nakamura2013,Hardhienata2011}.

To have an effective Maritime Domain Awareness (MDA), a number of
information sources are available, even though most maritime
infrastructure still lack a thorough response to the current
necessities~\cite{Mantzouris}. Most ship monitoring is currently
performed using maritime radars, vessel patrolling, and by ground
based Automatic Identification System
(AIS)~\cite{Narheim2008,Orr2013}. A significant part of the time is
spent in identifying unknown targets, collected from a variety of
Intelligence, Surveillance, and Reconnaissance (ISR) sensors. An
efficient process to perform this sorting, and to correlate
information with other ISR data, is critical. A possibility, suggested
by many, is the combination of AIS with SAR imagery. In some cases
this has already been initiated, for instance, by operations of the
Norwegian Coast Guard~\cite{Eriksen2010a}.

AIS is a ship-to-ship and ship-to-shore system used primarily to avoid
collisions, and to provide MDA and traffic
control~\cite{Orr2013,Pranajaya2010a}. It is mandatory for all
vessels with more than $300\,\unit{tonnes}$ and all passenger ships in
international waters, as also for ships with more than $500\,\unit{tonnes}$
in most national waters~\cite{Eriksen2010a}.

AIS works by sending a VHF signal providing the Maritime Mobile
Service Identity (i.e. position, heading, time, rate of turn, and
cargo)~\cite{Bedard2007}. It has a typical range between
$50\,\unit{km}$ and $100\,\unit{km}$, making it only practical
nearshore and in maritime choke points. However, due to this limited
range, the system can be simpler, using a self-organised time division
multiple access scheme (i.e. each transmitter within $100\,\unit{km}$
can self-organise and transmit its information without interfering
with messages sent from other ships in the same
cell)~\cite{Pranajaya2010a}.

When an AIS receiver is placed in space, a global view of maritime
traffic, with a number of applications can be achieved. These include
not only traffic awareness (with capabilities for national security
tasks and shipping), but also for search and rescue, and environmental
studies. In the fourth revision of the recommendations for the
system, AIS has been suggested as a means to promote ship-to-space
communication~\cite{Eriksen2010a}. However, in doing so, signal
jamming problems arise, due to the wider FOV (field of view) of a space-based platform
(normally more than the $100\,\unit{km}$), specially in high traffic
areas~\cite{Eriksen2010a}. Furthermore, because the signal has to
travel a longer distance (than the normal maximum $100\,\unit{km}$) it
is weaker. Another problem is the high relative velocity of the
spacecraft, which induces Doppler
shifts~\cite{Nakamura2013}. Moreover, the ionosphere induces a Faraday
rotation of the polarisation plane of the signals (dependent on its
frequency), decreasing the signal power received by the satellite (due
to the discrepancy between the signal and the antennas), a problem
also present with other remote sensing measurements.

Several spacecraft have carried AIS systems, and much work in this
area is still being performed~\cite{Takai2013}. For some, the literature is not
detailed, besides some general data and the launch dates. For instance,
\textbf{Triton 1} and 2 are two examples of 3U CubeSat spacecraft,
from the UK, that intend to test advanced AIS
receivers~\cite{Triton_GunterSpace}. \textbf{TianTuo 1} is a Chinese
nano satellite ($9.3\,\unit{kg}$) from the National University of
Defense Technology (NUDT) that also performs tracking of AIS signals,
in addition to other experiments~\cite{TT1_GunterSpace}. Two others
with less information available are \textbf{Perseus-M 1} and 2. These
spacecraft were built by Canopus Systems US, and follow the 6U
CubeSat standard~\cite{PerseusM_GunterSpace,PerseusM_Wiki}.
Finally, another example is the \textbf{DX 1} (Dauria Experimental 1), a
$27\,\unit{kg}$ spacecraft, with a size of
40$\,\times\,$40$\,\times\,$30 cm, built by Dauria
Aerospace~\cite{DX1_GunterSpace}. Besides testing technology it also
carries an AIS receiver.


\paragraph{\textbf{ZACube-2}} The Cape Peninsula University of
Technology (CPUT) created a programme to develop nano satellites in
South Africa, of which the \textbf{ZACube-2} will be the second
spacecraft~\cite{Villiers2015}. Although still not launched, this
spacecraft will serve as a technology demonstrator including a
software defined radio (SDR), which will be used for tracking AIS
signals, and a medium resolution imager, to perform ocean colour and
fire monitoring.


\paragraph{\textbf{M3MSat}} The Canadian Department of National
Defence (DND) wanted to expand the range of AIS farther than the
maritime inner zone, and integrate AIS and ISR data to produce an
improved Recognized Maritime Picture~\cite{Bedard2007}. Building upon
the experience of a previous mission, the Near Earth Object
Surveillance Satellite, a new satellite is scheduled to be launched in
2016, the \textbf{M3MSat} (Maritime Monitoring and Messaging
Microsatellite). The main mission objectives are: to monitor ship AIS
signals, utilised by the Canadian government and Exactearth (a
commercial venture for tracking AIS data); serve as a platform
demonstrator; and establish a flight heritage~\cite{Bedard2007}.


\paragraph{\textbf{AAUSat}} The Department of Electronic Systems of
Aalborg University (AAU) created an educational programme where
students could have access to many aspects of satellite design and
development. From this programme three satellites were designed for
tracking vessels with AIS systems, the \textbf{AAUSat3}, 4 and
5~\cite{AAUSAT_Page}. Of these, \textbf{AAUSat3} has concluded its
mission, \textbf{AAUSat5} was launched from the International Space
Station (ISS) in October 2015, whereas \textbf{AAUSat4} is still being
tested. With each spacecraft a new version of the AIS system is
flown.


\paragraph{\textbf{LambdaSat}} A group of international students in
Silicon Valley, California, have developed a spacecraft with three
main objectives: space qualification of graphene under direct solar
radiation and space exposure; demonstration of an AIS system; and
space qualification of three-fault tolerant
spacecraft~\cite{Mantzouris}.
The SmallSat was deployed in March 2015 from the
ISS~\cite{LambdaSat_EOPortal}. About 30 days later, the mission was
concluded and the spacecraft re-entered the atmosphere in May 2015.


\paragraph{\textbf{LAPAN-A2}} The Indonesian space agency (LAPAN)
propelled by the training of its own engineers in the Technical
University of Berlin, during the design and building of
\textbf{LAPAN-TUBSAT} or \textbf{A1}, developed the \textbf{LAPAN-A2}~\cite{LAPANA2_EOPortal}.
It was launched in 2015, for Earth
observation, disaster mitigation, and implementation of an AIS
system~\cite{Hardhienata2011}.


\paragraph{\textbf{AISat}} The Automatic Identification System
Satellite (\textbf{AISat}) was designed by the German Aerospace Center (DLR)
to monitor AIS signals. Launched in 2014, it is currently
operational. In particular, the Institute of Space Systems is
responsible for adapting the Clavis bus (an adaptation of the CubeSat platform by DLR)
to this mission, while the Institute of Composite Structures
and Adaptive Systems developed the helical antenna for the AIS
payload~\cite{Block2013}. Two companies (The Sch\"{u}tze Company and
Joachims) and the Bremen University of Applied Sciences were also
partners in the project. The difference between this mission and
others described here, is in the use of a helical antenna.


\paragraph{\textbf{Aeneas}} The \textbf{Aeneas} nano satellite was
designed to track cargo containers (equipped with a $1\,\unit{W}$ Wifi
type transceiver) around the world~\cite{Aherne2011}. It was developed
by students of the University of Southern California, Space
Engineering Research Center in partnership with iControl Inc. (the
primary payload provider) and was launched in
2012~\cite{UCSDatabase2015}. The orbit is an ellipse with a periapsis
of about $480\,\unit{km}$.


\paragraph{\textbf{VesselSat}} LuxSpace Sarl owns and operates two
micro satellites, the \textbf{VesselSat 1} (launched in 2011) and 2
(in 2012), to monitor maritime traffic with a space-based AIS
system~\cite{VesselSat_EOPortal}. Both continue to be
operational. These spacecraft were built upon the experience gained in
flying the \textbf{Rubin 7}, 8 and 9 non-separable payloads, flown
before on the upper stage of the Polar Satellite Launch Vehicle~\cite{Rubin9_GunterSpace}.


\paragraph{\textbf{SDS-4}} The Japan Aerospace Exploration Agency
(JAXA) created the Small Demonstration Satellite programme in 2006, in
order to test the next generation space technologies and, at the same
time, to create a standard $50\,\unit{kg}$ bus for future
missions~\cite{Nakamura2013}. Within this programme, the Small
Demonstration Satellite 4 (\textbf{SDS-4}) is the second satellite,
and one of its objectives, apart from demonstrating high performance
and small bus technology, is to validate a space-based AIS
system~\cite{Takai2013}. Launched in 2012, it remains operational,
even after completing its designed life time~\cite{SDS4_EOPortal}.


\paragraph{\textbf{AISSat-1}} Launched in 2010 (to a $630\,\unit{km}$
altitude Sun-synchronous polar orbit), \textbf{AISSat-1} is the first
dedicated satellite build for space-based monitoring of AIS signals by
Norway, in partnership with the University of Toronto, Institute for
Aerospace Studies/Space Flight Laboratory
(UTIAS/SFL)~\cite{Eriksen2010a,Narheim2008}. The objective is to
enhance MDA in Norwegian waters, which amounts to more than 2 million
square kilometres. Due to continuing success of the \textbf{AISSat-1} mission
(still operational in late 2015), two more spacecraft, the
\textbf{AISSat-2} and 3, where built (\textbf{AISSat-2} was launched in
2014, and \textbf{AISSat-3} is expected to be launched in 2016).


\paragraph{\textbf{NTS/CanX-6}} In 2007/2008 cooperation between the
University of Toronto, Institute for Aerospace Studies/Space Flight
Laboratory (UTIAS/SFL) and COM DEV Ltd. designed, produced and
launched a nano satellite, in less than seven months, to perform
experiments in the reception of AIS signals, the Nanosatellite
Tracking Ships (\textbf{NTS})~\cite{Pranajaya2010a}. While the
platform is based on SFL \textbf{CanX-2} nano satellite and the
Generic Nanosatellite Bus (GNB), the AIS payload was developed by COM
DEV. A top-down approach was performed, based on the mission
requirements (lifetime, data throughput, attitude, schedule, and
resources), balanced with a bottom-up analysis, which accounted for
hardware limitations (on-board memory, downlink rate, power, volume,
and others), maturity and readiness. Even though the spacecraft is
still in orbit, and keeps regular contact with the ground stations, it
is no longer considered operational.

\subsection{Constellations}
\label{subsec:Constellations}

Constellations can be considered as a category of their own. They are primarily used
to combine several observations of the same target, performed by
different platforms, and to obtain scientific data that could not be
acquired using a single spacecraft. Another usage concept for a
constellations is to achieve a permanent global coverage, like the GPS
constellation, or \textbf{Iridium} for communications. An example of such a constellation,
and of particularly relevance for this work, is the
\textbf{exactView} discussed below.


\paragraph{\textbf{SOCON}} The \textbf{SOCON} (Sustained Ocean
Observation from Nanosatellites) project, intends to develop the
\textbf{SeaHawk} 3U CubeSats spacecraft~\cite{SeaHawk_GunterSpace}.
Two prototypes are expected to be launched in 2017, and will act as
forerunners for a constellation of ten SmallSats. The
objective is to measuring ocean colour, using HawkEye Ocean Colour
Sensors. The associated cost is expected to be about eight times less,
with a resolution from seven to 15 times better, than its
predecessor, the single mini \textbf{OrbView-2} satellite (previously
named SeaStar).


\paragraph{\textbf{CYGNSS}} Each \textbf{CYGNSS} (Cyclone Global
Navigation Satellite System) satellite has a mass of $25\,\unit{kg}$,
and the constellation has eight identical satellites weighting in
total $200\,\unit{kg}$~\cite{CYGNSS_EOPortal}. This constellation,
expected to be launched in 2016 on a single launch vehicle, will serve
to relate ocean surface properties (retrieved from reflected GPS
signals), atmospheric conditions, radiation and convective dynamics,
with tropical cyclone formation~\cite{Rose2014}.


\paragraph{\textbf{exactView}} The \textbf{exactView}
spacecrafts/payloads are part of the exactEarth company constellation
of currently seven AIS systems~\cite{ExactEarth2015}. With this
constellation exactEarth expects to have a global revisit time (gap
between subsequent detection of individual ships) of around 90
minutes, giving their customers a comprehensive view of traffic in an
area of interest.

The first satellite was the experimental \textbf{NTS (EV0)}, discussed
above, which is considered by exactEarth to be retired. In 2011, came
the \textbf{EV2}, an AIS payload aboard the \textbf{ResourceSat 2} satellite, and
\textbf{EV6} and \textbf{5R} satellites (which were named \textbf{AprizeSat-6}
and 7) produced by SpaceQuest and transferred to exactEarth.
The fifth satellite, launched in 2012, is
\textbf{exactView1} (\textbf{EV1}). \textbf{AprizeSat 8} was commissioned and
became \textbf{EV12} in 2013. More recently, in 2014, \textbf{EV11}
and 13 (\textbf{AprizeSat 9} and 10) started operations. Three more
AIS systems are expected to be added to the constellation, the
\textbf{EV9} spacecraft (launched in September 2015), the \textbf{EV8} (which is
part of the payload of the Spanish \textbf{Paz} satellite), and the \textbf{M3MSat}
satellite discussed above (considered by exactEarth to be
\textbf{EV7})~\cite{exactEarth_constellation}.

The AprizeSat constellation goes beyond the 6 to 10 that passed onto
exactEarth. About 64 satellites are planned, so as to build a global
system of data communication, tracking and monitoring of assets, and in
some cases AIS payloads~\cite{AprizeSat_GunterSpace}.


\section{SmallSat Features}
\label{sec:Components_Features}

For completeness, we now present some features of a typical
spacecraft, while highlighting some specifics that are of importance
for ocean observation SmallSats. These platforms share most of the
characteristics of large spacecraft, adapted to the microcosm of a
SmallSat, and thus the information gathered here was obtained
from~\cite{Selva2012,Fortescue2011,Larson1999,Ley2009,Macdonald2014}.

Any spacecraft design is based on a \emph{bus}, which includes all
subsystems essential to its operation and to support the
\emph{payload}, its useful mission specific component. Some authors
also consider the booster adapter as an element to be part of the
spacecraft~\cite{Larson1999}.
For the purpose of this paper, the payload is the assembly of hardware
(and software) that senses or interacts with the object under
observation, in this case, the Earth's oceans. This is the focus of
Section~\ref{sec:Sensors_Oceanography}.

The bus performs the crucial functions of carrying the payload to the
right orbit and maintaining it there, pointing the payload in the
right direction, providing a structure to support its mass,
stabilising its temperature, supplying power, communications and data
storage if necessary, and handling commands and telemetry. The bus
is the physical infrastructure on which everything else is mounted in
hardware, or run in software. These functions are usually divided into
seven subsystems:

\begin{enumerate}
\item Structure and Mechanisms;
\item Propulsion systems;
\item Attitude Determination \& Control System (or guidance, navigation \& control) -- ADCS;
\item Power (or electric power) system -- EPS;
\item Thermal (or environmental) control system;
\item Command \& Data Handling (or spacecraft processor) -- C\&DH;
\item Communications (or tracking, telemetry and command) -- Comms.
\end{enumerate}

\begin{figure*}[!hbt]
  \centering
	\includegraphics[width=1\textwidth]{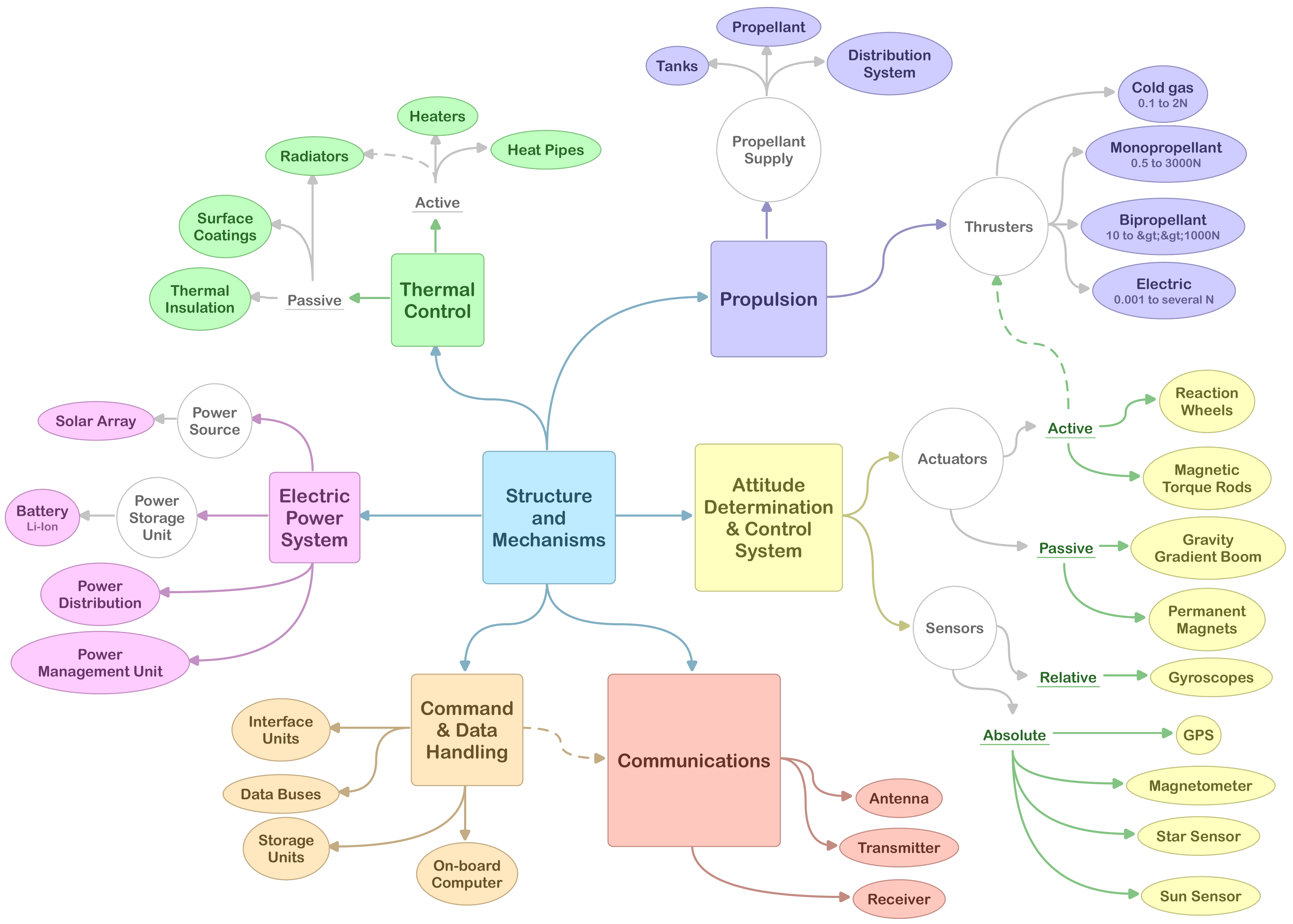}
  \caption{Schematic of a spacecraft subsystems (squares), with respective components (ellipses).}
  \label{fig:subsystems}
\end{figure*}

In many cases, the bus is a standard design adaptable with only minor
adjustments to a range of missions, each with a different
payload. This does not mean that any bus is adequate for any
payload. The aim of this section is to give a brief description of a
standard bus, while keeping the focus towards an oceanography oriented
SmallSat. In each part a brief description of the subsystem is
provided together with some limitations to be dealt with when
considering micro and nano satellites. Some of the subsystems are
crucial for oceanographic SmallSats, as their features will limit
scientific objectives. A schematic of these subsystems, with their
essential components, is shown in Fig.~\ref{fig:subsystems}.

\subsection{Structure and Mechanisms}
\label{subsec:Structures}

The structure and mechanisms subsystem is the skeleton of the
spacecraft, determining the overall configuration. It carries and
supports all the equipment, and protects (during launch) the
components to be deployed in orbit.
The structural architecture is designed based on the attitude control
requirements, and typically its mass range between 10 to 15\% of the
dry mass of the spacecraft.

For many SmallSats, the basic shape of the satellite is a cuboid (for
3-axis stabilised spacecraft), with usually five lightweight sides and
a more massive structure which supports the link to the launch
vehicle.
The spacecraft components (including payload) are linked to the inner
side of the exterior panels. Conversely, CubeSats have four pillars
(trusses or stiff rods) at the corners that support the structure, and
face plates (closed or open).

The University of Surrey, which is at the forefront of designing small
satellite platforms, also uses the CubeSat modular
approach~\cite{Fortescue2011}. Thus each subsystem is machined in
identical boxes, which are stacked on top of each other, and held
together by tie-rods, to form the main body, and to where other
instruments and solar panels are mounted onto.

Although in most missions the structure adapts to the payloads that
will be installed, there are some that work the other way around, i.e.
structure dimensions, and internal configuration, constrain the
payloads the satellite can carry. Payload restrictions is more evident
in CubeSats, since there is a standard for dimensions (and mass) to be
followed.

\subsection{Propulsion Systems}
\label{subsec:Propulsion}

A propulsion system serves mostly to change or to fix the orbit of a
spacecraft, including de-orbit manoeuvres at end of life, counteract
drag forces, and sometimes to control its attitude and angular
momentum. The most significant parameters of a propulsion system are
the total impulse, and thruster characteristics (number, orientation, and
thrust levels).
Other important parameters are total mass, power demand, reliability
and mission lifetime.

Common propulsion methods cover, in ascending order of performance:
cold gas, which can supply from $0.1\,\unit{N}$ to $2\,\unit{N}$;
monopropellant, with a capacity between $0.5\,\unit{N}$ and
$3000\,\unit{N}$; bipropellant, yielding $10\,\unit{N}$ to much more
than $1000\,\unit{N}$; and finally electric, with forces of just
$0.001\,\unit{N}$ to several N.
Although more efficient, electric propulsion requires a larger power
supply, than chemical propulsion, to produce thrust. Usually
solar panels are used, for Earth orbiting missions. This has an inherent
increase in mass (about $40\,\unit{W/kg}$).
 
Even though a propulsion system is included in most spacecrafts,
especially if in geosynchronous orbit (where station-keeping manoeuvres
are common), the simplest SmallSats \emph{do not} have any thrust
capability. Nevertheless, it has been installed in spacecraft as
small as CubeSats, based on either chemical or electrical systems.
An example, is the miniature cold gas system installed in the
\textbf{SNAP-1}, a $6.5\,\unit{kg}$ nano spacecraft.

\subsection{Attitude Determination \& Control System}
\label{subsec:ADCS}

The spacecraft's angular orientation (the direction it is pointing to)
is controlled and sensed by the attitude determination and control
subsystem (ADCS or ACS). In some cases, it also determines the
satellite's orbit (and it may even control it if a propulsion system
is present, as discussed in section~\ref{subsec:Propulsion}). Some
references treat orbit determination and control as a different
subsystem; we treat them together in this paper.

Characteristics and performance of the ADCS is determined by the
mission, and in particular, payload requirements. These affect the
accuracy (the difference between real-time and \textit{post facto}
attitude), the stability (the efficiency of keeping attitude rates),
and the agility (the time needed to change between two desired
attitudes). Other drivers include mass, power, cost, lifetime,
reliability, and redundancy~\cite{Macdonald2014}.

Some of the simplest spacecraft do not have any form of ADCS
subsystem, and the attitude will continuously change (one of the
oldest examples is the \textbf{Sputnik 1} itself). They can only have
payloads that are omnidirectional, like antennas, and all facets need
to have solar panels if powered by solar power. In other cases,
specially if the mass requirements are stringent, some SmallSats
employ passive control methods~\cite{Larson1999}. More complex
systems have controllers, actuators, and/or propulsion subsystems (to
actively control attitude, velocity, or angular momentum), and sensing
instruments to determine the attitude and orbit position~\cite{Ley2009}. 

There are several methods to determine the 3-axis attitude of a
spacecraft but, typically, data of different sensors are combined.
Common sensors include: star and sun sensors; magnetometers;
gyroscopes; and GPS.

Attitude control can be achieved either passively or actively.
Most common actuators are: permanent magnets; a gravity gradient boom;
thrusters; magnetic torque rods; and reaction wheels

For CubeSats, sensing accuracies of less than 2\textdegree\ have been
achieved (which can be translated to a ground uncertainty of
$14\,\unit{km}$ for a $400\,\unit{km}$ altitude orbit), while control
accuracies are less than 5\textdegree~\cite{Selva2012}. These
accuracies have to be improved for some oceanographic missions,
although the estimated 0.02\textdegree\ of future CubeSats (a ground uncertainty of
$140\,\unit{m}$) will certainly be enough for most applications. Sun
sensors with magnetometers are the preferred solution, combined with
passive or active magnetic control. Reaction wheels have also been
tested but are uncommon. Furthermore, advanced miniature GPS
navigation devices have already been used in nano
satellites for orbit determination.

A typical set of ADCS actuators for micro satellites, is a pack of
reaction wheels with magnetic torque rods, as these are necessary for
wheel unloading/desaturation, i.e. deceleration of the
wheels. Nevertheless, some passive control using permanent magnets has
also been used~\cite{Fortescue2011}. Usual sensors are a
magnetometer, a gyroscope, sun sensors (using most times complementary
metal–oxide–semiconductor, CMOS, detectors),
and a GPS. Star trackers are not as usual, due to size, power and
performance constraints, although there are some cases we have found
in our literature survey.

\subsection{Electric Power System}
\label{subsec:Power}

The electric power system (EPS) is another critical subsystem, of
which micro and nano satellites have to take the maximum advantage of.
The system needs to be autonomous and maintain the power supply
whatever the failure conditions~\cite{Ley2009}. An EPS provides
electric power for all other subsystems of the spacecraft, which
influences its size. Not only nominal power requirements are
important, but so is peak power consumption, and the orbit of the
satellite (for some types of energy sources).
Therefore, the EPS sizing limits the payloads carried by a spacecraft.

The basic system consist of a power source, a power storage unit,
power distribution, and a power management unit (including conversion,
conditioning, and charge and discharge).
Due to degradation of the power source and storage, system dimensions
must be made with end-of-life figures. Nevertheless, low mass and cost
is always the primary criteria
for SmallSats.

For Earth orbiting satellites, solar arrays equipped with photovoltaic
cells are the most common power source.
These are usually body-mounted, and/or in deployable rigid arrays,
using multi junction GaAs triple junction cells.

With a solar array for power source, which is dependent on solar
irradiation to generate power, the power storage unit takes special
relevance. Furthermore, it will also support the bus when peak power
is required (higher than what the solar array can supply). The typical
means to storage energy is a battery. If the spacecraft is to survive
more than a few weeks of mission life, batteries with charge
capabilities are needed, usually with Li-ion (Lithium-ion), NiCd
(nickel–cadmium), and NiMH sources.

\subsection{Thermal Systems}
\label{subsec:Thermal}

One problem that is sometimes not dealt with appropriately, is thermal
control, which can make a spacecraft inoperable. The thermal
subsystem is responsible to ensure that all components are kept within
the temperature range needed for optimal operation, which in-turn
determines its size.

A balance between heat loss and solar radiation received can be
achieved passively, through physical arrangement of equipment, thermal
insulation (e.g. thermal blankets) and surface coatings (e.g. paint).
However, passive control may not be enough, and thus active techniques
have to be implemented. These include heaters and heat
pipes. Radiators, which have surfaces with high emissivity and low
absorption, are also occasionally used actively or
passively~\cite{Macdonald2014}. For micro and nano satellites, passive
solutions are usually preferred due to mass saving and small space and
size. In particular, heat sinks and optical tapes are common options
for CubeSats~\cite{Selva2012}. Some active control (e.g. using joule
heating on the battery) has already been used. Nevertheless, when
photodiodes are present on a payload, active temperature control has
to be employed (in most cases), since they have stringent thermal
requirements.

\subsection{Command \& Data Handling}
\label{subsec:C_DH}

Responsibility for distributing commands controlling other subsystems,
handling sequenced or programmed events, and accumulation, storage and
housekeeping for payload data, rests on the command and data handling
system (C\&DH). This subsystem is closely linked to the communication subsystem,
to be discussed in section~\ref{subsec:Comm}. In particular, data
rates influences the system parameters, together with data
volume~\cite{Larson1999}. Other aspects to be addressed are
performance, reliability, self-healing (i.e. the capacity to treat
failures or anomalies), fault tolerance, the space environment
requirements (e.g. radiation and temperature), and power and size (which
is directly proportional to spacecraft complexity).
Common units are a central processor, data buses, interface and
storage units.

Usually, micro and nano satellites have a single on-board processor
which controls all aspects of the spacecraft. Another trend is the
increasing use of FPGA's (field-programmable gate array integrated
circuits). Furthermore, to have high data processing capabilities, at
the risk of less durability due to space radiation, COTS micro
controllers are being increasingly used on micro and nano satellites,
including commonly available ARM and PIC controllers.

\subsection{Communication Systems}
\label{subsec:Comm}

The communications subsystem assumes special relevance for micro and
nano satellites, since limitations of power and size will limit the
spacecraft's capability to transmit data. This subsystem links the
spacecraft to the ground or in some special cases to other spacecraft.
This is a critical system, since without communications a mission can
be considered lost in practice, and hence, redundancy and as much as
possible near spherical (4$\pi$) coverage, must be implemented.

A receiver, a transmitter, and an antenna (which can be directional,
hemispheric, or omnidirectional) are the basic components of a
communication system. Each of these elements is selected and sized
taking into account the desirable data rates, error rate allowance,
communication path length, and radio frequency. The transmitting
power, coding and modulation are also important factors. There are
COTS components for each of these elements, with specific reliability
characteristics. The communication bandwidth needed for a SmallSat is
related to the power the spacecraft can generate, and correlated to the
altitude at which the platform operates. Higher the altitude, less
friction with the atmosphere, resulting in lower orbit decay rate, but it comes
with a higher power requirement to communicate.

This survey of SmallSats confirms a trend found by other surveys,
namely that 75\% of CubeSats use UHF (ultra-high frequency)
communications, with rates reaching $9.6\,\unit{kbps}$, and 10\% VHF
(very high frequency), with similar data rates~\cite{Selva2012}. Only
15\% have S-band systems, with maximum rates of $256\,\unit{kbps}$
(although in our survey the \textbf{NTS} SmallSat has the maximum
rate, $32\,\unit{kbps}$). The power required for communications can
reach up to $1\,\unit{W}$, although it is usually smaller. These small
rates have a great impact on the science that can be performed.

For micro satellites S-band communications are the most common, and
there are also cases of C, X and Ku bands. Typical values for data
rates are around $1\,\unit{Mbps}$, for S-band, but can reach
$2.5\,\unit{Mbps}$ with the X-band (\textbf{RISESat}), or even an impressive
$30\,\unit{Mbps}$ on the \textbf{QSat-EOS} spacecraft (using the
Ku-band). Nevertheless, most SmallSats keep to UHF, although as more
power becomes available, higher data rates can also be achieved; for
instance the \textbf{VesselSat} has a UHF data rate of
$512\,\unit{kbps}$.

\subsection{Regulating SmallSat Communications}
\label{subsec:Reg_Comm}

With the growing interest of an increasing number of nations in
developing their indigenous space capacity, there is a growing concern
in the space community about the lack of adherence of CubeSat missions
to international laws, regulations and procedures, as stated in the
Prague declaration on Small Satellite Regulation and Communication
Systems~\cite{InternationalTelecommunicationUnion2015}.

Developers and members of the SmallSat community have been asked, by
the United Nations agency (the International Telecommunication Union,
ITU), to re-evaluate current frequency notification procedures for
registering SmallSats. Most users however, consider this procedure
complex and unduly laborious, and not suitable for such low-cost
missions. Consequently, the ITU has been asked to examine the
procedures for notifying space networks, and consider possible
modifications to enable SmallSat deployment and operation. Ideally,
this should be done taking into account the short development and
mission time, and unique orbital characteristics of such platforms.

In March 2015, preliminary results of a study showed that there are no
specific characteristics that are relevant from a frequency management
perspective~\cite{Ubbels2015}. Nevertheless, in April 2015, the UN
Office for Outer Space Affairs (OOSA) and the ITU has issued a best
practice guide~\cite{Subcommittee2015}. It summaries the laws and
regulations that are applicable for launching small satellites, and
include the following:

\begin{itemize}

\item Notification and recording of the radio frequencies used by a
  satellite at the ITU by its national authority;

\item Consideration of space debris citation measures in the design
  and operation of SmallSats (to guarantee their re-entry in less
  than 25 years after the completion of the mission);

\item After launch, registration of the spacecraft with the
  Secretary-General of the United Nations by its national authority.

\end{itemize}

Furthermore, SmallSat missions are under the Liability Convention
resolution 2777 (XXVI), which establishes potential liabilities if a
collision happens~\cite{UnitedNationsOfficeforOuterSpaceAffairs1971}.
Consequently, a variety of countries are approving national space
laws, including an obligatorily insurance to cover liabilities derived
by collisions during the platforms life.

Finally, in November 2015, the Provisional Final Acts of the World
Radiocommunciation Conference recommends that
frequencies used for telemetry, tracking and command
communications, by satellites with missions lasting less than three
years, should be preferably within the following ranges: $150.05$ -- $174\,\unit{MHz}$
and $400.15$ -- $420\,\unit{MHz}$~\cite{Int2015}.


\section{Sensors for Oceanography}
\label{sec:Sensors_Oceanography}

The use of remote sensing data has been an integral part of
oceanography~\cite{Robinson2004}. Typically, sea surface temperature,
ocean colour, winds, and sea-state have been the traditional foci of
such data collection. More recently, SAR has been become a significant asset to
both scientific and security related purposes. In oceanography, often
remote sensing data allows for feeding synthetic ocean models (e.g
ROMS~\cite{ROMS}), which in turn allows for predications, and can be
used for sampling via manned or robotic
assets~\cite{Das-2010-637,bloomML}.

Large spacecraft typically from NASA and the National Oceanographic and Atmospheric
Administration (NOAA) or ESA
have been a continuous source of data over a substantial span of
time~\cite{OrbView2_EOPortal,TerraSARX_EOPortal}. More recent trends
show that SmallSats are being preferred over large conventional
satellite missions~\cite{Fortescue2011}. Although this has not yet
evolved to the use of micro or nano spacecrafts, we believe that
SmallSats will end up achieving more cost-effective ways to observe
the global ocean.

In this section we describe the primary oceanographic features
and the detectors that have been used in remote sensing. The main
source for this discussion
are~\cite{Selva2012,Robinson2010,Robinson2004,CEOS_Database,EOPortal,Chelton2001a,Martin2014}.

\subsection{Ocean colour}
\label{subsec:Ocean_colour}

The coastal upper water-column is often a region of high primary
productivity especially due to the presence of phytoplankton with
chlorophyll~\cite{Martin2014}. Through measurement of ocean colour,
one can infer the concentrations of sediments, organic material and
phytoplankton, whose quantities and type influence colour.
Anthropomorphic input primarily from sewage, fertiliser run off or
commercial dumping, provides a nutrient base for organisms in the
coastal ocean. Therefore, this influences productivity and in turn
phytoplankton concentration, as determined by chlorophyll
density. Consequently, remote sensing provides a wide-scale view in a
cogent manner.

Other ocean phenomena that can be observed using
ocean colour as a proxy include mesoscale eddies (circular currents on
the ocean spanning $10\,\unit{km}$ to $500\,\unit{km}$ in diameter and
that persist from a few days to months), fronts (boundaries between
distinct water masses), upwelling (when deep cold water rises to the
surface) and internal waves.

Ocean colour observations using spacecraft began in 1978 with the Coastal
Zone Color Scanner (\textbf{NIMBUS-7})~\cite{Martin2014}. However,
only in 1996 other missions were launched: the Japanese Ocean Color
and Temperature Sensor (\textbf{ADEOS-1}); and the German Modular
Optical Scanner (on the \textbf{IRS-P3}). In that same year, the
International Ocean-Colour Coordinating Group was established to
support ocean colour technology and studies, backed by the Committee
on Earth Observation Satellites (CEOS)~\cite{IOCCG_Web}. It has
brought together data providers, space agencies, and users, namely
scientists and managers. Moreover, it sets the standards for
calibration and validation of measurements.

A typical ocean colour sensor samples with a high spectral resolution
in several bands, including ultraviolet, visible and near
infrared. Two types of sensors exist: a multispectral radiometer,
which has a limited number of narrow wavelengths to capture the
structure of the incoming light; and an imaging spectrometer, which
samples across the spectrum with a defined spectral resolution,
generating substantially more data. The combination of different
spectrum bands helps to distinguish the colour origin, and overcome
some atmospheric interference. Nevertheless, precise measurements are
difficult to perform due to shortcomings of the instruments (e.g. it
demands a cloud free observation), and accuracies are of about 50\%.

Large spacecraft with mass well above the small satellite limit of
$1000\,\unit{kg}$ have flown with ocean colour sensors
(e.g. \textbf{EnviSat}, \textbf{Aqua}, and \textbf{COMS-1}).
There are only two examples of mini satellites, both carrying
radiometers, the \textbf{OceanSat-2} of the Indian Space Research
Organization (ISRO), which has
$970\,\unit{kg}$~\cite{OceanSat2_EOPortal}, and the Chinese
\textbf{Haiyang-1B}, with $443\,\unit{kg}$~\cite{HY1B_EOPortal}.
Resolution, swath and other instrument data of some ocean colour
sensors are shown in Table~\ref{tab:Ocean_colour}. Nevertheless, it
is perfectly possible to include a ocean colour sensor in a CubeSat,
and two examples are described below.

As noted earlier, \textbf{SeaHawk} (of the \textbf{SOCON} constellation) and
the \textbf{ZACube-2}, which are scheduled to be launched in 2016,
will measure ocean colour using nano satellites. In particular, the
\textbf{SOCON} project (a 3U CubeSat) will fly the HawkEye Ocean Colour
instrument. This is expected to have a ground
resolution of $75\,\unit{m}$ per pixel (for a $540\,\unit{km}$ orbit),
and a total of 4096$\,\times\,$10000 pixels (each image), for a ground
swath of
300$\,\times\,750\,\unit{km}$~\cite{Seahawk_WebInfo}. Measuring eight
bands (similar to VIS and NIR of the SeaWiFS instrument), with eight
linear CCDs, will generate a total of $4.6\,\unit{Gb}$, to be
aggregated to $655\,\unit{Mb}$ that must be
downlinked~\cite{Seahawk_WebInfo}.

\iftoggle{paperlayout}{
\begin{table*}[!t]
  \centering
  \caption{Ocean colour instruments}
	\label{tab:Ocean_colour}
	\footnotesize
		\begin{threeparttable}
    \begin{tabular}{ccccc}
    \noalign{\smallskip}\hline\noalign{\smallskip}
    Sensor  						& Figures of merit\tnote{a}	& COCTS\ \cite{HY1B_EOPortal}	& OCM-2\ \cite{OceanSat2_EOPortal} 	& HawkEye\ \cite{Seahawk_WebInfo} \\
		(Satellite)					& (Past/Future)							& (HY-1B) 										& (OceanSat-2) 											& (SeaHawk) \\
    \noalign{\smallskip}\hline\noalign{\smallskip}
    Spatial Resolution [km] & 1.5/0.7								& 1.1													& 0.36  														& 0.15 - 0.075 \\
    Swath [km] 					& 1328/1474									& 2800												& 1420  														& 750 \\
    Wavelength [$\mu$m]	& 0.404 - 5.42							& 0.402 - 0.885								& 0.404 - 0.885 										& - \\
    Mass [kg]  					& 122/242										& 50													& 78    														& - \\
    Power [W]  					& 99/206 										& 29.3												& 134 															& - \\
    \noalign{\smallskip}\hline\noalign{\smallskip}
    \end{tabular}
		\begin{tablenotes}
			\item [a] Average figures of merit of decommissioned and future sensors~\cite{Robinson2004,EOPortal,IOCCG_Web}.\\
		\end{tablenotes}
		\end{threeparttable}
\end{table*}
}{
\begin{table}[!t]
  \centering
  \caption{Ocean colour instruments}
	\label{tab:Ocean_colour}
		\begin{threeparttable}
    \begin{tabular}{ccccc}
    \noalign{\smallskip}\hline\noalign{\smallskip}
    Sensor  						& Figures of merit\tnote{a}	& COCTS\ \cite{HY1B_EOPortal}	& OCM-2\ \cite{OceanSat2_EOPortal} 	& HawkEye\ \cite{Seahawk_WebInfo} \\
		(Satellite)					& (Past/Future)							& (HY-1B) 										& (OceanSat-2) 											& (SeaHawk) \\
    \noalign{\smallskip}\hline\noalign{\smallskip}
    Spatial Resolution [km] & 1.5/0.7								& 1.1													& 0.36  														& 0.15 - 0.075 \\
    Swath [km] 					& 1328/1474									& 2800												& 1420  														& 750 \\
    Wavelength [$\mu$m]	& 0.404 - 5.42							& 0.402 - 0.885								& 0.404 - 0.885 										& - \\
    Mass [kg]  					& 122/242										& 50													& 78    														& - \\
    Power [W]  					& 99/206 										& 29.3												& 134 															& - \\
    \noalign{\smallskip}\hline\noalign{\smallskip}
    \end{tabular}
		\begin{tablenotes}
			\item [a] Average figures of merit of decommissioned and future sensors~\cite{Robinson2004,EOPortal,IOCCG_Web}.\\
		\end{tablenotes}
		\end{threeparttable}
\end{table}
}

\subsection{Ocean altimetry}
\label{subsec:Ocean_altimetry}

Altimetry has been used to retrieve surface topography (including sea
level and wave height), ocean currents, and bathymetry (submarine
topography). Additionally, it is one of the most reliable ways to
observe mesoscale eddies, detected by small displacements of the sea
surface elevation,
and scaler wind speed.
Thus, data acquired from altimeters have applications not only to
oceanography, but also to worldwide weather and climate patterns.

The most common instruments for altimetry are nadir pointing (looking
vertically downward) radar altimeters, sampling along the ground
track. The instrument works by emitting short regular pulses, and
recording the travel time, magnitude and shape of the returned signal.
From the travel time one can get the range from the satellite to the
sea surface, to which corrections due to the atmosphere and
ionospheric free electrons, sea state effects, and instrument
calibrations need to be considered.
Combining this with other measurements, including precise gravity
fields (from the model created by GRACE~\cite{GRACE_EOPortal}), and
orbit position using other instruments, the range is converted into
the height of the sea surface relative to the reference ellipsoid. Sea
surface roughness, for length scales of the radar wavelength (that is
from a few millimetres to centimetres), and sea surface height
variability, on the instrument footprint, can also be retrieved. These
yield estimates for wave height and wind speed.

Accuracies of $1\,\unit{cm}$ can be achieved, with measurement precisions
of $6\times10^{-11}\,\unit{s}$,
but most often are in the $2\,\unit{cm}$ to $3\,\unit{cm}$ range.
However, the obtained accuracies are only for open ocean, since
coastal waters induce other effects~\cite{Vignudelli2006}. Altimeter
constellations are deemed important, since they bring an increase in
temporal resolution, and some ocean phenomenon can only be perceived
if subject to an almost continuous observation. At the same time, a
higher revisit time represents an increase in spatial coverage and a
finer spatial sampling grid, for a single altimetry sensor.
Equally, sun synchronous orbits should be avoided, because of the
errors associated with solar tidal effects.

The most successful missions are compact satellites (with low drag
resistance),
which are equipped with supporting sensors, to measure the necessary
corrections and determine orbital position. These supporting sensors
consist of: a radiometer -- for atmospheric corrections; the Doppler
Orbitography and Radiopositioning Integrated by Satellite (DORIS) -- a
precise orbit determination instrument using ground beacons spread
over the world; and a Laser Retroreflector Array -- which provides a
reference target for satellite laser ranging measurements. Although
DORIS has been widely used, many spacecraft are replacing it by a
Global Navigation Satellite Systems (GNSS) instrument (or adding to
DORIS a GNSS sensor). Another measure that may be important is the
inertial position.

Past missions equipped with altimeters oscillated between large
spacecraft with more than $2000\,\unit{kg}$ mass (\textbf{SeaSat},
\textbf{ERS-1}, and \textbf{ERS-2}), and mini satellites with mass
around $500\,\unit{kg}$ (\textbf{GEOS-3} and
\textbf{GEOSAT})~\cite{EOPortal}. Examples of recent altimetry
missions, and which use dual frequency (C and Ku-band) with
increased accuracy,
are: \textbf{TOPEX/Poseidon} -- with $2388\,\unit{kg}$ and active
between 1992 and 2005;
\textbf{Jason-1} -- active from 2001 to 2013, with a mass of
$500\,\unit{kg}$~\cite{Jason1_EOPortal}; and finally \textbf{EnviSat}
-- one of the heaviest so far, with a mass of $8140\,\unit{kg}$, that
was operational between 2003 and 2012, and was equipped with several
instruments~\cite{EnviSat_EOPortal}. Currently, three mini satellites
are dedicated to altimetry, namely \textbf{SARAL} (with Ka-band for
which ionospheric delay corrections are substantially
reduced)~\cite{SARAL_EOPortal}, \textbf{Jason-2} (C and
Ku-bands)~\cite{Jason2_EOPortal}, and \textbf{CryoSat-2} (Ku-band with
two antennas)~\cite{CryoSat2_EOPortal}.

Typically, an altimetry mission requires a payload mass of about
$100\,\unit{kg}$, a power consumption of $150\,\unit{W}$, and one
antenna with $1.1\,\unit{m}$ in diameter (averaging the values of the
three currently flying mini satellite missions). Features from past,
current and future altimeter instruments are presented in
Table~\ref{tab:Ocean_altimetry}.

From the figures presented above, it appears to be challenging for
altimetry to be performed with a SmallSat platform (nano satellites
are obviously out of the question), if the same accuracy is
required. Nevertheless, there has been some effort to miniaturise
Ka-band radar altimeters.
A study of a constellation of twelve small satellites (with an
expected mass of less than $150\,\unit{kg}$), \textbf{GENDER}, was
made in 1999, but the objective was only to obtain significant wave
height and wind speed (and not the full sea surface
topography)~\cite{Zheng1999,DaSilvaCuriel1999}. In 2008, two
institutes linked to the previous altimetry missions made a proposal
of a micro satellite, with $45\,\unit{kg}$ and a power budget of
$70\,\unit{W}$, with errors of around
$5.6\,\unit{cm}$~\cite{Richard2008}. Another proposal, disclosed in
2012, was based on a 6U CubeSat ($12\,\unit{kg}$)~\cite{Stacy2012}.

Nevertheless, even with higher errors, more spatial measurements have
the potential to increase the return on investment~\cite{Mroczek2015}.
Consequently, there are studies for nano satellites with altimeters. Making
a scaling exercise from the \textbf{Jason-2} altimeter
($70\,\unit{W}$) to a nano satellite instrument consuming
$1\,\unit{W}$, the error would increase from $2\,\unit{cm}$ to a worst
case scenario of $16.7\,\unit{cm}$~\cite{Mroczek2015}. However, this
is just for the altimeter and does not account for errors induced by
orbit position determination, due to the lack of supporting
sensors. The atmospheric corrections can nevertheless be introduced
via suitable modelling (which have accuracies of about 1 to
$3\,\unit{cm}$).

\iftoggle{paperlayout}{
\begin{table*}[!t]
  \centering
  \caption{Decommissioned, current and future altimeter parameters and supporting sensors}
	\label{tab:Ocean_altimetry}
	\footnotesize
		\begin{threeparttable}
    \begin{tabular}{ccccc}
    \noalign{\smallskip}\hline\noalign{\smallskip}
    Sensor 									& Figures of merit\tnote{a}	& Poseidon-3\tnote{b}\ \ \cite{Jason2_EOPortal} & AltiKa\tnote{b}\ \ \cite{SARAL_EOPortal} 	& Optimised\ \cite{Richard2008} \\
    (Satellite)	 						& (Past/Future)      				& (Jason-2) 																		& (SARAL) 																	& Micro Satellite \\
    \noalign{\smallskip}\hline\noalign{\smallskip}
		Spatial Resolution [km]	& -/8.5 										& -     																				& -     																		& - \\
    Swath [km] 							& 13/68 										& 30    																				& 8     																		& - \\
    Bands 									& Ku/C or Ka 								& Ku/C  																				& Ka    																		& Ku \\
    Accuracy [cm] 					& 6.1/2.4 									& 2     																				& 1.8   																		& 2.5 \\
    Mass [kg] 							& 108/142 									& 70    																				& 40    																		& 13 \\
    Power [W] 							& 134/329 									& 78    																				& 75    																		& 24 \\
    Antenna [m] 						& 1.2/2.5 									& 1.2   																				& 1     																		& 0.4 \\
    Supporting Sensors 			& \multicolumn{3}{c}{Radiometer, GPS, DORIS, LRA\tnote{c}} 																													& GPS \\
    \noalign{\smallskip}\hline\noalign{\smallskip}
    \end{tabular}
		\begin{tablenotes}
			\item [a] Average figures of merit of decommissioned and future sensors~\cite{Robinson2004,EOPortal,Chelton2001a}.
			\item [b] These are representative of current capacities of mini satellites.
			\item [c] SARAL does not have GPS and AltiKa also works as a radiometer.
		\end{tablenotes}
		\end{threeparttable}
\end{table*}
}{
\begin{table}[!t]
  \centering
  \caption{Decommissioned, current and future altimeter parameters and supporting sensors}
	\label{tab:Ocean_altimetry}
		\begin{threeparttable}
    \begin{tabular}{ccccc}
    \noalign{\smallskip}\hline\noalign{\smallskip}
    Sensor 									& Figures of merit\tnote{a}	& Poseidon-3\tnote{b}\ \ \cite{Jason2_EOPortal} & AltiKa\tnote{b}\ \ \cite{SARAL_EOPortal} 	& Optimised\ \cite{Richard2008} \\
    (Satellite)	 						& (Past/Future)      				& (Jason-2) 																		& (SARAL) 																	& Micro Satellite \\
    \noalign{\smallskip}\hline\noalign{\smallskip}
		Spatial Resolution [km]	& -/8.5 										& -     																				& -     																		& - \\
    Swath [km] 							& 13/68 										& 30    																				& 8     																		& - \\
    Bands 									& Ku/C or Ka 								& Ku/C  																				& Ka    																		& Ku \\
    Accuracy [cm] 					& 6.1/2.4 									& 2     																				& 1.8   																		& 2.5 \\
    Mass [kg] 							& 108/142 									& 70    																				& 40    																		& 13 \\
    Power [W] 							& 134/329 									& 78    																				& 75    																		& 24 \\
    Antenna [m] 						& 1.2/2.5 									& 1.2   																				& 1     																		& 0.4 \\
    Supporting Sensors 			& \multicolumn{3}{c}{Radiometer, GPS, DORIS, LRA\tnote{c}} 																													& GPS \\
    \noalign{\smallskip}\hline\noalign{\smallskip}
    \end{tabular}
		\begin{tablenotes}
			\item [a] Average figures of merit of decommissioned and future sensors~\cite{Robinson2004,EOPortal,Chelton2001a}.
			\item [b] These are representative of current capacities of mini satellites.
			\item [c] SARAL does not have GPS and AltiKa also works as a radiometer.
		\end{tablenotes}
		\end{threeparttable}
\end{table}
}

\subsection{Ocean surface winds}
\label{subsec:Ocean_winds}
 
Ocean surface wind measurements have been performed since 1978, and
have implications for atmospheric, ocean surface waves and circulation
models.
These measurements have also enhanced marine weather forecasting and
climate prediction which have a direct impact on offshore oil
operations, ship movement and routing.

Surface winds can be derived from surface waves distribution, and the
most effective bands for wind speed retrieval are the microwave C, X,
and Ku bands.
Nevertheless, measurement accuracies are always dependent on the wind
speed. There are two main methods used in satellites to retrieve
surface wind speed and direction: passive microwave radiometers; and
active radar instruments (including scatterometers and SAR). In most
instruments there is a trade-off between spatial resolution and swath.

An effective sensor is the scatterometer, which is also the simplest
type of radar for remote sensing.
They work by emitting microwaves at incidence angles between 20\textdegree\ and
70\textdegree, and measuring the average backscatter of the signals
reflected by the same patch of sea within a wide field of view.
Each area has to be viewed several times, either from different
directions or at different polarisations.
Therefore, there are two types of scatterometers: fixed vertical fan
beams pointing in a single direction, thus requiring several antennae;
and focussed beam, which perform circular scans.
Typically, some corrections have to be implemented to obtain valid data,
such as noise, atmospheric attenuation, and ambiguities. The
measurement then goes through a selected model to retrieve wind speed
and direction.

Conversely, SAR measurements can only be used if the wind direction is
previously known (so scatterometer backscatter models can be applied),
since SAR views the ocean from only one direction, unlike
scatterometers. We provide a detailed view on SAR in
section~\ref{subsec:SAR}.

An alternative to active instruments, is a radiometer, which measures
the microwave radiation emitted by the sea, passively. This radiation
has a dependence on sea surface shape and orientation, besides water
temperature and dielectric properties, allowing for wind speed
retrieval.
However, to retrieve wind direction, the radiometer that uses only
vertical and horizontal polarisation must look twice at the same area.
Whereas, fully polarimetric radiometers i.e. those that measure all four
Stokes parameters (vertical and horizontal polarisation components,
and the corresponding real and imaginary parts of the correlation
between both polarisation components of the electric field) can
retrieve the wind vector with a single look under some conditions (as
the four parameters describe the properties of an arbitrarily polarised
electromagnetic wave).
The basic components of a typical mechanically scanning radiometer
are: a single parabolic reflector; a cluster of feed horns; and
microwave detectors and the corresponding gimbal.
An alternative to the gimbal is to use an array of small antennae, and
combine the inputs electrically, an approach used in the \textbf{SMOS}
mission~\cite{SMOS_EOPortal}.

There were, and still are, many missions using these sensors to
retrieve ocean surface winds. Many missions equipped with altimeters
had launch mass higher than $2000\,\unit{kg}$ (e.g. \textbf{HY-2A},
\textbf{RADARSAT-2}, and \textbf{TRMM}),
but some are of the SmallSat class:
\textbf{QuikSCAT}~\cite{QuikSCAT_EOPortal} and
\textbf{OceanSat-2}~\cite{OceanSat2_EOPortal}, both using active
rotating scatterometers; and \textbf{DMSP}~\cite{DMSP_EOPortal},
\textbf{Jason 1} and 2~\cite{Jason1_EOPortal,Jason2_EOPortal}, and
\textbf{Coriolis}~\cite{Coriolis_EOPortal}, all with passive rotating
radiometers. Furthermore, the altimeter on \textbf{SARAL} also works
as a radiometer~\cite{SARAL_EOPortal}. Relevant characteristics of
scatterometers and radiometers flown or currently active are presented
in Table~\ref{tab:Ocean_winds}.

Based on previous missions, an average radiometer would have a mass of
$118\,\unit{kg}$, a power consumption of $122\,\unit{W}$, and a
reflector of $116\,\unit{cm}$. Conversely, for the scatterometer, this
is a $238\,\unit{kg}$ instrument, consuming $221\,\unit{W}$, and using
a $120\,\unit{cm}$ antenna.

Consequently, a scatterometer would be extremely difficult (if not
impossible) to fit on a nano satellite platform, due to mass and power
requirements as well as mechanical constraints. Miniature radiometers
have been suggested as viable options to be used on CubeSats, although
with significant doubt on the utility of the measurements. For micro
spacecrafts, the average radiometer with the values presented above,
would be more feasible, even though the mechanical needs could be
difficult to accomplish. Without some in depth research, in principle,
a scatterometer would not fit a SmallSat, since it has double of the
mass of the platform.

\iftoggle{paperlayout}{
\begin{table*}[htb]
  \centering
  \caption{Decommissioned and active instruments to measure ocean surface winds}
	\label{tab:Ocean_winds}
	\footnotesize
    \begin{threeparttable}
    \begin{tabular}{cccc}
    \noalign{\smallskip}\hline\noalign{\smallskip}
    Sensor 														& Figures of merit\tnote{a}	& SeaWinds\tnote{b}\ \ \cite{QuikSCAT_EOPortal} & WindSat\tnote{c}\ \ \cite{Coriolis_EOPortal} \\
    (Satellite) 											& (Past)      							& (QuikScat) 																		& (Coriolis) \\
    \noalign{\smallskip}\hline\noalign{\smallskip}
    Spatial Resolution [km]						& 50    										& 50    																				& 25 \\
    Swath [km] 												& 1000  										& 1800  																				& 1000 \\
    Wind Accuracy (Wind speed) [m/s]	& 2 (4 -- 20)								& 2 (3 -- 20)																		& 2 (3 -- 25) \\
    Wind Direction Accuracy [deg] 		& 20    										& 20    																				& 20 \\
    Bands 														& Ku/C  										& Ku   	 																				& C/X/K/Ka \\
    Frequency [GHz] 									& 13.9/5.3 									& 13.4  																				& 6.8/10.7/18.7/23.8/37 \\
    Mass [kg] 												& 230   										& 205   																				& 341 \\
    Power [W] 												& 197   										& 250  	 																				& 350 \\
    Antenna [m] 											& 1.7   										& 1     																				& 1.8 \\
    \noalign{\smallskip}\hline\noalign{\smallskip}
    \end{tabular}
		\begin{tablenotes}
			\item [a] Average figures of merit of decommissioned sensors~\cite{Robinson2004,EOPortal,Martin2014}.\\
			\item [b] The only currently active scatterometer in a mini satellite.\\
			\item [c] An example of a currently active fully polarimetric radiometer in a mini satellite.
		\end{tablenotes}
		\end{threeparttable}
\end{table*}
}{
\begin{table}[htb]
  \centering
  \caption{Decommissioned and active instruments to measure ocean surface winds}
	\label{tab:Ocean_winds}
    \begin{threeparttable}
    \begin{tabular}{cccc}
    \noalign{\smallskip}\hline\noalign{\smallskip}
    Sensor 														& Figures of merit\tnote{a}	& SeaWinds\tnote{b}\ \ \cite{QuikSCAT_EOPortal} & WindSat\tnote{c}\ \ \cite{Coriolis_EOPortal} \\
    (Satellite) 											& (Past)      							& (QuikScat) 																		& (Coriolis) \\
    \noalign{\smallskip}\hline\noalign{\smallskip}
    Spatial Resolution [km]						& 50    										& 50    																				& 25 \\
    Swath [km] 												& 1000  										& 1800  																				& 1000 \\
    Wind Accuracy (Wind speed) [m/s]	& 2 (4 -- 20)								& 2 (3 -- 20)																		& 2 (3 -- 25) \\
    Wind Direction Accuracy [deg] 		& 20    										& 20    																				& 20 \\
    Bands 														& Ku/C  										& Ku   	 																				& C/X/K/Ka \\
    Frequency [GHz] 									& 13.9/5.3 									& 13.4  																				& 6.8/10.7/18.7/23.8/37 \\
    Mass [kg] 												& 230   										& 205   																				& 341 \\
    Power [W] 												& 197   										& 250  	 																				& 350 \\
    Antenna [m] 											& 1.7   										& 1     																				& 1.8 \\
    \noalign{\smallskip}\hline\noalign{\smallskip}
    \end{tabular}
		\begin{tablenotes}
			\item [a] Average figures of merit of decommissioned sensors~\cite{Robinson2004,EOPortal,Martin2014}.\\
			\item [b] The only currently active scatterometer in a mini satellite.\\
			\item [c] An example of a currently active fully polarimetric radiometer in a mini satellite.
		\end{tablenotes}
		\end{threeparttable}
\end{table}
}

\subsection{Sea surface temperature}
\label{subsec:Ocean_temperature}

The Global Climate Observing System has described sea surface
temperature (SST), as an essential climate variable (together with sea
surface salinity and ocean colour).
With SST, one can retrieve information on current systems, eddies,
jets, and upwelling regions, since these are closely linked to ocean
circulation. Furthermore, SST moderates the global climate system, and
influences atmospheric water vapour and heat fluxes, making it a
critical boundary condition for many atmospheric models.
Therefore, it is an essential part for global climate studies, such as
the El Ni\~{n}o Southern Oscillation cycle, and it supports local
climatology, including ship routing, and hurricane forecast. The
sensor's accuracy requirements for most of the studied phenomenon are
of at least $0.5\,\unit{K}$, with short revisit time.

Typical satellite sensors for SST are closely related to those used in
ocean colour, such as infrared systems, and for surface wind
retrieval, such as passive microwave systems. While infrared payloads
need a cloud free for observation, microwave instruments using
appropriate channels can perform measurements through no rain induced
cloud cover,
with necessary atmospheric corrections. However, the microwave data
cannot be used within $75\,\unit{km}$ of land, due to a reduction of
signal to noise ratio (the noise coming from ground communications
systems). The infrared sensor usually requires a black body of known
temperature for on-board calibration.
Although temperature difference requirements are possible with current
instruments, to achieve the needed revisit time (i.e. to have more readings of the temperature of a specific place), satellite data and
data collected \textit{in situ} (e.g. with buoys, AUVs and ships) is
usually combined.

Typically, many of the spacecraft performing these measurements have
mass over the $1000\,\unit{kg}$ limit of SmallSats
(e.g. \textbf{EnviSat} and \textbf{Terra}).
Within this limit there are only two mini satellites: \textbf{Meteosat-7} (with
$696\,\unit{kg}$)~\cite{Meteosat7_EOPortal} and \textbf{EO-1}
($572\,\unit{kg}$)~\cite{EO1_EOPortal}. The first, has a high
resolution 3-band (thermal infrared, water vapour absorption and
visible range bands) passive radiometer, spinning with the platform.
The normal data transmission rate of the sensor is about
$330\,\unit{kbps}$, but can reach $2.7\,\unit{Mbps}$ in burst mode.
\textbf{EO-1} has two instruments that can be used to retrieve SST:
the panchromatic and multi spectral imager ALI,
which has a data rate of $300\,\unit{Mbps}$;
and the hyper spectral imager Hyperion, which uses two spectrometers,
and creates multiple images per acquisition, each with
$75\,\unit{MB}$. Some figures of decommissioned, future and the two
current mini satellites is shown in Table~\ref{tab:Ocean_temperature}.

The described instruments could be adapted to fit a micro satellite,
although the amount of data that they generate would be challenging to
downlink. For CubeSats it is believed that microbolometers (an
infrared sensor) can be a viable option to measure sea surface
temperature.
However, there are no recorded SmallSat missions with such an instrument.

\iftoggle{paperlayout}{
\begin{table*}[!htb]
  \centering
  \caption{Decommissioned and currently active instruments to measure ocean surface temperature}
	\label{tab:Ocean_temperature}
	\footnotesize
    \begin{threeparttable}
    \begin{tabular}{ccccc}
    \noalign{\smallskip}\hline\noalign{\smallskip}
    Sensor 									& Figures of merit\tnote{a}	& MVIRI\ \ \cite{Meteosat7_EOPortal}	& ALI\ \ \cite{EO1_EOPortal}	& Hyperion\ \ \cite{EO1_EOPortal} \\
    (Satellite) 						& Past/Future 							& (Meteosat 1st) 											& \multicolumn{2}{c}{(EO-1)} \\
    \noalign{\smallskip}\hline\noalign{\smallskip}
		Spatial Resolution [km] & 15/0.9 										& 5     															& 0.03  											& 0.03 \\
    Swath [km] 							& 1030/2200 								& -     															& 37    											& 7.5 \\
    Accuracy [K] 						& 0.65/0.35 								& 1     															& -     											& - \\
    Wavelength [$\mu$m] 		& 0.6 - 12 									& 0.5 - 12.5 													& 0.48 - 2.35									& 0.4 - 2.5 \\
    Mass [kg] 							& 190/196 									& 63    															& 106   											& 49 \\
    Power [W] 							& 205/173 									& 17    															& 100   											& 78 \\
    \noalign{\smallskip}\hline\noalign{\smallskip}
    \end{tabular}
		\begin{tablenotes}
			\item [a] Average figures of merit of decommissioned and future sensors~\cite{Robinson2004,CEOS_Database,EOPortal}.
		\end{tablenotes}
		\end{threeparttable}
\end{table*}
}{
\begin{table}[!hbt]
  \centering
  \caption{Decommissioned and currently active instruments to measure ocean surface temperature}
	\label{tab:Ocean_temperature}
    \begin{threeparttable}
    \begin{tabular}{ccccc}
    \noalign{\smallskip}\hline\noalign{\smallskip}
    Sensor 									& Figures of merit\tnote{a}	& MVIRI\ \ \cite{Meteosat7_EOPortal}	& ALI\ \ \cite{EO1_EOPortal}	& Hyperion\ \ \cite{EO1_EOPortal} \\
    (Satellite) 						& Past/Future 							& (Meteosat 1st) 											& \multicolumn{2}{c}{(EO-1)} \\
    \noalign{\smallskip}\hline\noalign{\smallskip}
		Spatial Resolution [km] & 15/0.9 										& 5     															& 0.03  											& 0.03 \\
    Swath [km] 							& 1030/2200 								& -     															& 37    											& 7.5 \\
    Accuracy [K] 						& 0.65/0.35 								& 1     															& -     											& - \\
    Wavelength [$\mu$m] 		& 0.6 - 12 									& 0.5 - 12.5 													& 0.48 - 2.35									& 0.4 - 2.5 \\
    Mass [kg] 							& 190/196 									& 63    															& 106   											& 49 \\
    Power [W] 							& 205/173 									& 17    															& 100   											& 78 \\
    \noalign{\smallskip}\hline\noalign{\smallskip}
    \end{tabular}
		\begin{tablenotes}
			\item [a] Average figures of merit of decommissioned and future sensors~\cite{Robinson2004,CEOS_Database,EOPortal}.
		\end{tablenotes}
		\end{threeparttable}
\end{table}
}

\subsection{Ocean salinity}
\label{subsec:Ocean_salinity}

Ocean surface state has numerous effects on the climate of the planet,
affecting the global hydrological cycle, on which differences in ocean
salinity concentrations can have large repercussions.
Together with temperature, salinity concentration has an impact on the
density and stability of surface water, influencing ocean mixing and
the water-mass formation processes, such as on the meridional
overturning circulation~\cite{CEOS_Database,Schmittner2013}. Salinity
measurements, therefore, would improve ocean modelling and
analysis.

Passive microwave radiometers are used to obtain readings for the
surface salinity from satellites. The most suitable frequency for this
is $1.41\,\unit{GHz}$, in the L-band. To acquire valid and relevant
information, salinity should be determined with at least
$0.2\,\unit{psu}$\footnote{Seawater of $35\,\unit{psu}$ (precision
salinity units), has a conductivity ratio of unity at 15 degrees
Centigrade at 1 atmosphere, for 32.4356 grams of KCl (potassium
chloride) per kilogram of solution.}.
Phenomena that require corrections include: sea surface roughness;
ionospheric effects (inducing Faraday rotation); and heavy rain.
Salinity can also be determined using brightness temperature (or
radiance of microwave radiation) measurements, if the sea surface
temperature is known. To accomplish that, the brightness temperature
measurements accuracy must be about $0.1\,\unit{K}$~\cite{Martin2014}.
Typically, moored buoys also perform salinity measurements, and can be
used to ground-truth satellite data.

Currently, two missions perform these measurements, the \textbf{SMOS}
(a $670\,\unit{kg}$ spacecraft)~\cite{SMOS_EOPortal}, and the
\textbf{SAC-D/Aquarius} Mission. \textbf{SMOS} is a SmallSat with only one
payload, the Microwave Imaging Radiometer using Aperture Synthesis,
which weighs $369\,\unit{kg}$ and has a power consumption of
$375\,\unit{W}$. This is a synthetic aperture radiometer, with 69
antenna/receiver modules distributed in a deployable three arm (each
with a length of $8\,\unit{m}$) structure, creating a sparsely
populated antenna. The signals of each receiver is electronically
combined, forming a two dimensional interferometer microwave image.

Currently, there are no radiometers that would fit on a nano satellite
platform, and studies conducted raise doubts on the validity of
measurements performed by a small instrument. Further research might
allow for fitting a radiometer in a micro satellite. Finally, as
stated above, there is a correlation between brightness temperature
measurements and salinity, which could lead to microbolometers (that
fit on micro or nano satellites) for measuring salinity. However, due
to the low spectral resolution, this has still to be demonstrated.

\subsection{GNSS Reflectometry}
\label{subsec:GNSS_Reflectometry}

As shown in the previous sections, most instruments used for
oceanography have characteristics that makes them difficult to
implement on micro and nano satellites. Additionally, many instruments
suffer from limitations due to the frequency band in which they
operate, which is often blocked by clouds and heavy rainfall
conditions. However, a promising method to perform some of these
measurements has been proposed, namely the use of Global Navigation
Satellite Systems Reflectometry (GNSS-R). Although limited for some
measurements, some of its advantages are the capability for operating
in all precipitating conditions, as well as the availability of GNSS
signals~\cite{Rose2014}. GNSS-R would work as a complement to other
instruments and not as a substitute, since the expected accuracy is
lower.

A smooth surface causes a specular point reflection of a signal
(i.e. a mirror like reflection where the incoming signal is reflected
in a single outgoing direction), whereas a rough surface causes a
signal scattering. Measurement of this scattering allows for
obtaining information of the surface, which for the case of the ocean
gives information on the sea and wave height, and indirectly, about
the near surface meteorological conditions. While most radars use this
principle, they have to actively produce a signal to be reflected. As
with radiometers, which passively measure the natural radiation
emitted by the surface, the idea behind GNSS-R is to use GPS signals
(or other navigation satellite constellations such as Galileo), to
create an image of the scattering cross section in time and frequency;
this is called the Delay Doppler Map~\cite{Rose2014}. This is
achieved by combining the GNSS signal coming directly from the GNSS
constellation, providing a coherent reference, as well as signals
reflected from the ocean. By analysing both, a number of ocean
parameters can be inferred with a centimetre scale
accuracy~\cite{Meerman2002,Rose2012}.

The idea for using GNSS signals to perform ocean scatterometry
(measuring ocean near-surface wind speed and direction) was proposed
in 1988~\cite{Rose2012}. In 1993, GNSS-R was also suggested as another
means for altimeter measurements~\cite{Meerman2002}. Moreover, ocean
roughness data acquired through GNSS-R contributes to other areas of
research, such as for ocean salinity, ocean circulation and sea ice
studies (as well as some land studies)~\cite{Rose2014,Foti2015}.

A mission was proposed in 2002 to fit a GNSS-R in a $12\,\unit{kg}$
micro satellite (and a $4.8\,\unit{W}$ orbit average
power)~\cite{Meerman2002}. The payload in this proposal was a GPS
receiver (with a nadir pointing and another zenith pointing antenna)
and a solid-state data recorder. There were also studies of a GNSS-R
instrument to assist radiometry for sea surface salinity. Even though
it does not fit a CubeSat, it could work on a micro
satellite~\cite{Selva2012,Camps2008a}. In 2003, the GNSS sensing
technique was tested in the \textbf{UK-DMC} satellite, even though
only $20\,\unit{s}$ of data could be acquired each
time~\cite{Rose2012,TechDemoSat1_EOPortal}. In 2012, a proposal was
made to employ GNSS for altimetry using twelve nano satellites in a
non-polar orbit with the GNSS Ocean Wind and Altimetry
mission~\cite{Rose2012}. This constellation would provide data with
an accuracy between $7\,\unit{cm}$ and $20\,\unit{cm}$, for ocean
topography, and more than $70\,\unit{m/s}$, for wind speed
measurements. A spatial resolution of about $30\,\unit{km}$, and
a revisit time of less than two days, is also expected.
Another proposed mission was a single $5.4\,\unit{kg}$ nano
satellite, with an average power consumption of $5.5\,\unit{W}$, and
an expected payload data volume of $13.8\,\unit{Mb/day}$, the
\textbf{$^3$Cat-2}~\cite{CarrenoLuengo2013}. In 2014, a more robust
and extensive test of this technique was performed with the GNSS-R
payload (with a mass of around $1.5\,\unit{kg}$ and approximately
$10\,\unit{W}$ power consumption) of the
\textbf{TechDemoSat-1}~\cite{Foti2015}. Finally, the \textbf{CYGNSS}
constellation (described in section~\ref{subsec:Constellations}) will also use GNSS-R to perform ocean surface readings, and is
expected to be launched in 2016~\cite{Rose2014}.

\subsection{The case of (and for) SAR}
\label{subsec:SAR}

Synthetic Aperture Radar is a radar imaging system which has
been increasingly used in a range of civil and security related
applications. In particular, it has been critical for environmental
monitoring and planning, especially in the context of maritime domain
awareness (MDA) and marine spatial planning~\cite{Pitz2010}.
Furthermore, SAR can provide valuable information about the oceans,
including surface and internal waves, shoals, sea ice, rainfall, and
location of man-made platforms and large marine
mammals~\cite{pineda2015whales,da2015internal}. One of the areas
where SAR has been extremely successful is surface winds on coastal
zones. Equally interesting is recent work in using SAR imagery to
estimate local bathymetry, which has enormous implication for
military applications worldwide~\cite{mishra2014}. Although, Doppler readings for
currents has been used, it continues to be challenging from an
algorithmic perspective~\cite{Robinson2004}.

Radar imaging systems (like SAR) use an antenna to send a sequence of
short microwave pulses to the surface of the ocean. Their
reflection is measured, usually using the same antenna, in the
intervals when no pulses are being sent. The principle that
distinguishes SAR from other radar systems, is that by combining
several individual readings, while in view of a target, it synthesises
a single image with high resolution. Since they are microwave based,
the signal can penetrate clouds without major energy loss, allowing
the system to work with cloud cover. A more detailed discussion of
the working principle of SAR is found in the appendix.

The first spacecraft carrying SAR, NASA's \textbf{SEASAT}, was
launched in 1978. It was followed by other missions from several
countries~\cite{McCandless2004}. Table~\ref{tab:SAR} shows the major
spacecraft equipped with SAR, which are currently active. All are
spacecraft with low, near polar, sun synchronous orbits with a period
of about 100 minutes. The SAR instrument is usually turned on for
approximately 10 minutes in each orbit due to its high energy demand.

Currently it is challenging to fulfil the user needs of SAR. That
has led ESA to focus the use of their \textbf{Sentinel-1} spacecraft
to view some regions of the globe they believe are of interest, since
there is considerable lack of global coverage. Other SAR-based
spacecraft still try to be open to scientist's requests for specific
area observations. \textbf{TerraSAR-X}, for instance, makes some of the
images available to the public. \textbf{RADARSAT 2} has better
accuracy and resolution, but its images are less freely accessible.

\iftoggle{paperlayout}{
\begin{table*}[!ht]
  \centering
  \caption{Earth observing spacecraft with SAR payloads.}
  \label{tab:SAR}
	\footnotesize
	\begin{threeparttable}
    \begin{tabular}{lcccccc}
      \noalign{\smallskip}\hline\noalign{\smallskip}
      Satellite 									& RADARSAT 2 					& TSX/TDX 								& COSMO-SkyMed 					& KOMPSAT-5 			& Sentinel 1\\
																	& \cite{Staples2002} 	& \cite{Steinbrecher2014} & \cite{DiLazzaro2008} 	& \cite{Lee2010} 	& \cite{Snoeij2009} \\
      \noalign{\smallskip}\hline\noalign{\smallskip}
      Launch date 								& 2007 								& 2007 										& 2007 									& 2013 						& 2014 \\
			Sat. Mass [kg]\tnote{a}			& 2200								& 1230/1340 							& 1700 									&	1400						& 2300 \\
      Sat. Altitude [km]\tnote{a}	& 798 								& 514 										& 620 									& 550 						& 693 \\
			SAR Mass [kg]\tnote{a}			&	750									& 394											& -											& 520							& 945 \\
			Peak Power [W]\tnote{a}	  	& 1650								& 2260										& -											& 1700						& 4075 \\
      Swath [km] 									& 20 -- 500						& 30 -- 260 							& 30 -- 200 						& 30 -- 100 			& 80 -- 400 \\
      Resolution [m] 							& 3 -- 100 						& 1.7 -- 10 							& 1 -- 100 							& 1 -- 20 				& 5 -- 40 \\
      System Band									& C 									& X 											& X 										& X 							& C \\
      \noalign{\smallskip}\hline\noalign{\smallskip}
    \end{tabular}
		\begin{tablenotes}
			\item [a] Figures from~\cite{EOPortal}.
		\end{tablenotes}
		\end{threeparttable}
\end{table*}
}{
\begin{table}[!ht]
  \centering
  \caption{Earth observing spacecraft with SAR payloads.}
  \label{tab:SAR}
  \begin{threeparttable}
    \begin{tabular}{lcccccc}
      \noalign{\smallskip}\hline\noalign{\smallskip}
      Satellite 									& RADARSAT 2 					& TSX/TDX 								& COSMO-SkyMed 					& KOMPSAT-5 			& Sentinel 1\\
																	& \cite{Staples2002} 	& \cite{Steinbrecher2014} & \cite{DiLazzaro2008} 	& \cite{Lee2010} 	& \cite{Snoeij2009} \\
      \noalign{\smallskip}\hline\noalign{\smallskip}
      Launch date 								& 2007 								& 2007 										& 2007 									& 2013 						& 2014 \\
			Sat. Mass [kg]\tnote{a}			& 2200								& 1230/1340 							& 1700 									&	1400						& 2300 \\
      Sat. Altitude [km]\tnote{a}	& 798 								& 514 										& 620 									& 550 						& 693 \\
			SAR Mass [kg]\tnote{a}			&	750									& 394											& -											& 520							& 945 \\
			Peak Power [W]\tnote{a}	  	& 1650								& 2260										& -											& 1700						& 4075 \\
      Swath [km] 									& 20 -- 500						& 30 -- 260 							& 30 -- 200 						& 30 -- 100 			& 80 -- 400 \\
      Resolution [m] 							& 3 -- 100 						& 1.7 -- 10 							& 1 -- 100 							& 1 -- 20 				& 5 -- 40 \\
      System Band									& C 									& X 											& X 										& X 							& C \\
      \noalign{\smallskip}\hline\noalign{\smallskip}
    \end{tabular}
		\begin{tablenotes}
			\item [a] Figures from~\cite{EOPortal}.
		\end{tablenotes}
		\end{threeparttable}
\end{table}
}

\subsection{The case for SmallSat SAR}

The use of SAR in SmallSats poses some significant challenges. The spacecraft in
Table~\ref{tab:SAR} have masses ranging from about $1200\,\unit{kg}$ to
$2300\,\unit{kg}$. The SAR antennas used until recently have a length
of 12 meters, although more recent instruments use shorter antennas of around 5
meters. Israel's military satellite \textbf{TecSAR} uses a smaller
deployable antenna, however, with an umbrella shape, with a mass of
less than $0.5\,\unit{kg}$~\cite{Sharay2005}. Another challenge is how
to generate the power needed to obtain a reasonable signal-to-noise
ratio.

That said, we believe SAR is a potential SmallSat sensor, which can provide a
significant boost to the use of SmallSats for oceanography and for
studying the climate, while providing information to policy makers and
for environmental monitoring. Coupled with \textit{in situ}
measurements that robotic platforms can make
(Section~\ref{sec:robots}), we believe that SAR can provide novel ways
to observe the changing oceans systematically and persistently. In
particular, for seaward facing nations such as Portugal, Spain and
Norway, situational awareness in their territorial waters has been
problematic, especially given significant increases in ship-borne
traffic. With increasing access to the polar regions, for instance,
where search and rescue on-demand is challenging, MDA is critical to
mitigate environmental pollution and loss of life. Portugal, for
instance, should its ongoing application to the United Nations for the
Extension of the Continental
Shelf~\footnote{\url{http://www.emepc.pt/} in Portuguese.} prove to be
successful, will need to cover a vast swath of the mid-Atlantic, over
and beyond the Azorean archipelago, to monitor shipping. As a
reflection of the complexity, should the extension be granted, it is
estimated that upwards of 65\% of all maritime traffic to the European
Union will pass through waters under Portuguese jurisdiction. Ship or
aircraft-based surveillance and monitoring is clearly not
sustainable. Although they can be augmented with robotic vehicles, the
sheer size of the area requires a space-borne all-weather radar asset
that is affordable. In addition, the Azores and Madeira archipelagos
provide a bellwether to understand our changing climate and, for
instance, the impact it may have under extreme weather conditions. We
believe that SAR on SmallSats is a technology that is viable for
exploitation within the next 3 to 5 year time frame\footnote{We note that
the Norwegian Defense Establishment (FFI) is well advanced towards
building a mini-SAR for use in UAVs (personal communication).
We believe the
transition to SmallSat operation could be a viable follow-up.}.


\section{Discussion and Conclusions}
\label{sec:discuss}

The advantages of small satellites have been made clear throughout
this paper: smaller costs and
shorter development cycles allow for a greater flexibility.
Indeed, there is a strong correlation between a
satellite's size and its cost and time-to-launch. The shorter
development cycle is enabled, in part, by the use of
commercial-off-the-shelf components for some of the main
subsystems. This allows not only for the exploitation of a
greater number of mission opportunities, but also for the
incorporation of newer technologies, either more advanced, cheaper, or
both, in an agile development cycle.

There is another important benefit from the development of small
satellites. The fact that they have been, and predictably will primarily be, built
by small teams, makes university environments ideal for such
projects. The cost structure allows for such a viable mode of
operation. For example, at the University of Vigo we have recently
launched our third SmallSat, opening a huge potential for future
science oriented missions.

In addition to the lower costs and shorter design cycle, there is a
critical pedagogical component. SmallSats can be designed, built,
tested and potentially flown in space within the life-cycle of a
graduate or undergraduate student's academic programme. And they have
been operated predominantly by students when in flight. The life-cycle
of the satellite provides a systematic view of how such complex
systems can be built for young researchers, providing key life-long
skills for a new generation of space engineers and scientists.

There are, however, still some important limitations for SmallSats and
the range of their application, in the context of ocean
observation. The examples listed in the survey on
Section~\ref{sec:Survey}, fall into three main categories: Ocean
Imaging, Data Relay and Tracking.

This last category is the most numerous, especially with the usage of
space-borne ship-tracking systems like AIS, of great interest for many
private and public institutions. The strategic and economical benefits
are evident when considering the increase in ship traffic, due to the
growth in global trade, the potential opening of new waterways, not to
mention limitations and cost of land based solutions.

While not directly applicable to oceanography, some small satellites
carrying communications and data relay payloads can be useful in the
context of connecting sensors in the open ocean with land based
facilities. Furthermore, as noted in Section~\ref{sec:robots},
providing situational awareness and remote sensing data for
oceanographic field experiments and persistent observations is a great
asset.

It is also worth mentioning the significant potential of deploying
small satellite constellations. These can be used to increase the
capabilities, by combining different payloads that would otherwise be
combined in a larger satellite, or the surface coverage. Some examples
are already slated for launch as noted in Section~\ref{sec:Survey}.

The usage of imagers and other remote sensing payloads is still very
limited in SmallSats. Some remote sensing satellites are among the
heaviest and most complex ever launched. It is still unclear if and
which remote sensing instruments can be adapted for use in small
satellites, and when they will be available. Nevertheless, there is a
substantial interest in the oceanographic community in increasing the
availability and affordability of such instrumentation, with low(er)
cost satellite operations.

For example, the measurement of ocean colour is quite relevant in
coastal regions, where it can help to detect the presence and
concentration of phytoplankton blooms or plumes, with scientific and
economic ramifications. Section~\ref{sec:Sensors_Oceanography}
discusses other examples of payloads that would be useful for
oceanography.

The instrument that would, perhaps, have the greatest potential, would
be radar-based and specifically SAR, as discussed in
Section~\ref{subsec:SAR}. This is primarily due to the wide range of
applications that SAR has, from monitoring shipping, to ocean
dynamics, to providing information on bathymetry for civil and
military needs. Adding to the fact that SAR can work independent of
cloud cover. Although there are significant technical challenges to
overcome before a SAR can be installed in a SmallSat, the potential
scientific, economic and security related benefits are
substantial. This is especially true for countries with large
territorial waters under their jurisdiction, that might necessitate
the mobilisation of vast sea and airborne resources for effective
monitoring and surveillance. We argue that using SmallSats instead
could augment such needs with the use of radar based sensors and could
potentially be achievable within the next 5 to 10 years.


%
 \section*{Acknowledgements}
 The work of A. G. is supported by the Funda{\c c}\~ao para a Ci\^encia
 e a Tecnologia (Portuguese Agency for Research) fellowship
 PD/BD/113536/2015.  K. R. is supported by United States Office of
 Naval Research, ONR Grant \# N00014-14-1-0536. 

\section*{References}

\bibliographystyle{elsarticle-num}
\bibliography{smallsat,kanna-ref}

\newpage
\section*{Appendix A}
\label{sec:sar-math}

In this section, we briefly discuss the technicalities associated with
SAR, which we believe has yet to play a large and critical role in
space-based oceanography, for civil and security applications, with SmallSats.

A typical side looking radar aboard a satellite is shown in
Fig.~\ref{fig:SAR}, with the most relevant parameters. We represent
the orbit of the spacecraft, its projection on the ground, and two
extra directions, the range (perpendicular to the projection of the
orbit on the ground) and the azimuth (parallel to the orbit
projection), for convenience.

If the altitude of the satellite is $h$, its range, $R$, is the
distance from the spacecraft to the centre of the region on the ground
illuminated by the antenna. As the figure represents a side looking
radar, the imaging antenna points to the ground at an angle $\theta$
with the vertical, the incident angle. As $\theta > 0$, the swath
width, $S$, the width of the illuminated area in the range direction,
is increased. Furthermore, the antenna's length in the direction of
the spacecraft's motion, $D$, is usually bigger than the perpendicular
length, $d$, as shown. Since the angle of the beam is inversely
proportional to the length of the antenna, this also makes the
illuminated region wider in the range, and narrower in the azimuth
direction.

\iftoggle{paperlayout}{
\begin{figure}[!htb]
  \centering
  \includegraphics[width=1\columnwidth]{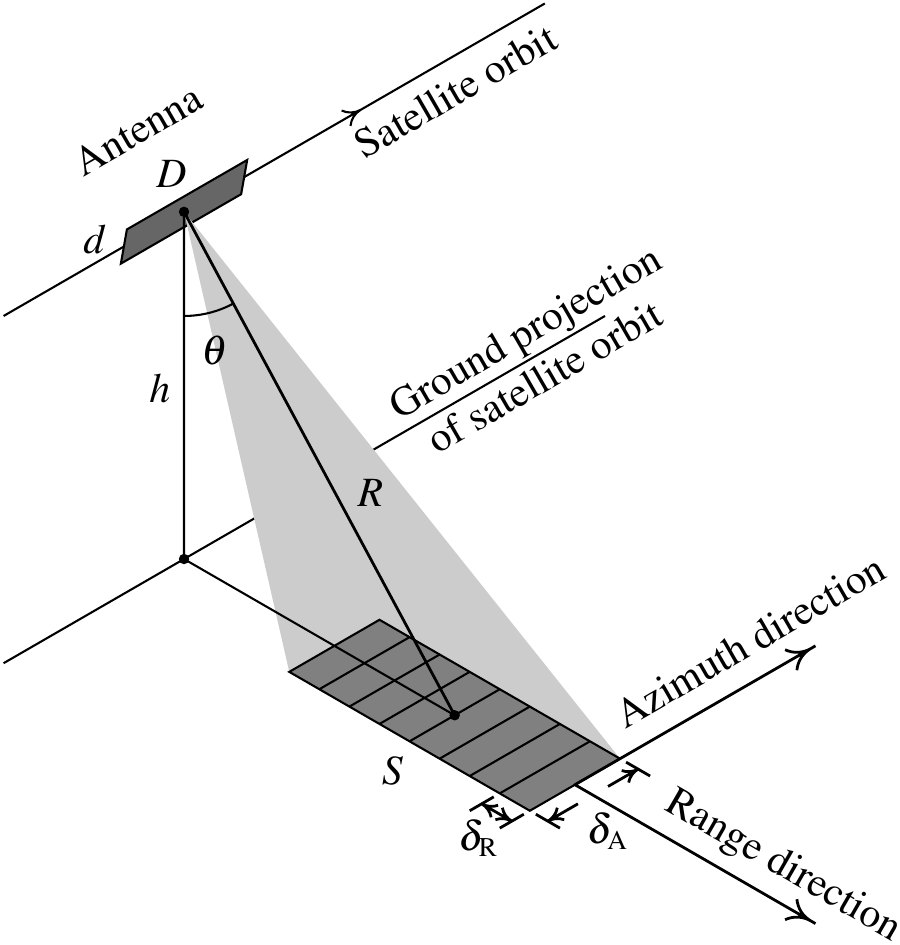}
  \caption{Illustration of a side looking imaging radar system on a space platform.}
  \label{fig:SAR}
\end{figure}
}{
\begin{figure}[!htb]
  \centering
  \includegraphics[width=0.5\textwidth]{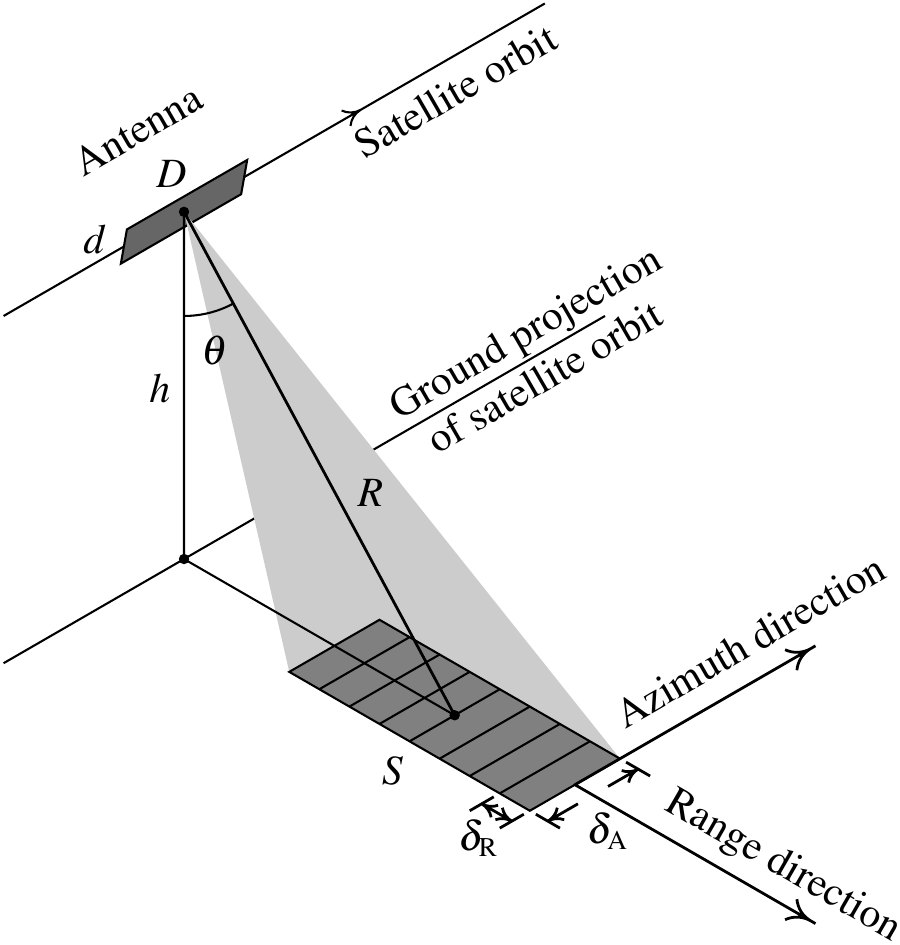}
  \caption{Illustration of a side looking imaging radar system on a space platform.}
  \label{fig:SAR}
\end{figure}
}

Consequently, different points along the range direction are
characterised by different time delays between the emitted pulse and
the return of the back-scattered signal to the antenna. A pulse
reflected from a point with a distance $R_i$, from the antenna, takes a time
$2\,R_i/c$ to get back, where $c$ is the speed of
light. To distinguish two different points, the difference between
their time delays must be larger than the time, $\tau$, that each pulse
lasts. Therefore, the resolution distance, $\delta_\mathrm{R}$, in the
range direction is approximately
\begin{equation}
  \label{eq:delta_R}
  \delta_\mathrm{R} \approx \frac{c\tau}{2}.
\end{equation}

For the altitude of the spacecraft, typically a few hundreds of
kilometres, very short pulses of a few nanoseconds are required to
attain a resolution of the order of tens of meters. However, to give
those short pulses the minimum energy to allow the reflected
pulse to be distinguishable from the background noise, the peak intensity of
the emitted pulse would need to be large. Thus, the required
intensity would burn the antenna. The problem can be solved by sending a
frequency-modulated (chirped) pulse, i.e. a pulse whose frequency
decreases with time. The reflected pulse detected in the antenna is
then filtered, delaying the higher frequencies. That shortens the time
span, $\tau$, of the detected pulses, which turn them equal to $1/B$,
where $B$ is the bandwidth (that is, the difference between the higher and
lower frequencies of the emitted pulse). The resolution in the range
direction then becomes
\begin{equation}
  \label{eq:resolution_R}
  \delta_\mathrm{R} \approx \frac{c}{2B}.
\end{equation}

The resolution in the azimuth direction, $\delta_\mathrm{A}$ in
Fig.~\ref{fig:SAR}, depends only on the capability of the antenna to
receive independent pulses from two points without interference, and
it is approximately equal to
\begin{equation}
  \label{eq:delta_A}
  \delta_\mathrm{A} \approx \frac{\lambda R}{D},
\end{equation}

\noindent{}where $\lambda$ is the wavelength of the signal used.

An important parameter in radar systems is the signal-to-noise ratio
(SNR). The expression for the SNR is usually called the radar
equation. Various radar equations are found in the
literature, specific to different types of
images~\cite{Cumming2005,EEE355}. A general expression can be written
as
\begin{equation}
  \label{eq:SNR}
  SNR = \frac{P_t\,G_t\,G_r\,\lambda^2\,A\,\sigma_0}{(4\pi)^3R^4\,k\,T\,B\,F_n\,L_s},
\end{equation}

\noindent{}where $P_t$ is the average transmitted power, $G_t$ and
$G_r$ are the gains of the transmitter and receiver antennas
respectively (often equal since a single antenna is usually used), $A$
is the area of the target and $\sigma_0$ its normalised cross section
(the fraction of the power of the radar radiation which is returned
by the target in the direction of the receiver). In the denominator,
we have Boltzmann's constant, $k$, the absolute temperature of the
radar system, $T$, the receiver noise figure, $F_n$, and $L_s$ that
accounts for the system losses. As before, $R$ is the range,
$\lambda$ the wavelength of the radar radiation, and $B$ is the
bandwidth of the emitted pulse.

For a typical radar, for instance, with a bandwidth of
$15\,\unit{MHz}$, a range resolution of $10\,\unit{m}$ can be
attained, and the emitted pulse can have a longer duration, $\tau$, of
the order of tens of micro seconds. The bandwidth cannot be increased
much further because it poses technical difficulties, and the
possibility of getting off the range of frequencies allowed for
microwave radars (without interfering with other telecommunication
devices)~\cite{Robinson2004}.

Conversely, a wide antenna, with $D=10\,\unit{m}$, is not capable to
achieve an azimuth resolution better than tens of meters. In fact,
for microwaves, in order to get a reasonable image resolution, an
antenna of several kilometres would be needed.

It is in this context, that SAR manages to outperform other radar systems.
By taking advantage of the motion of the antenna relative to the
target, a point in the ground reflects back the pulses to the antenna
during the time that the satellite illuminates it. Meanwhile, the
antenna moves a sizeable distance, receiving back-scattered pulses.
Combining all signals, it is as if the antenna
had the length of the distance that the satellite as moved while the
target is still in view. In other words, the effective length of a SAR
antenna, $D_\mathrm{SAR}$, is of the order of the size of the
illuminated region itself, yielding a much larger azimuth
resolution. To distinguish pulses coming from different points along
the azimuth direction, the Doppler shift of the received pulses is used~\cite{Kovaly1976}.

To fully exploit SAR features we should then consider the minimum area
of the antenna,
$A_{min} = \delta_\mathrm{R}\delta_\mathrm{A}/\sin\theta$, and the
``velocity'', $v = D/\tau$. Applying this to the the radar equation,
it follows the SAR equation
\begin{equation}
  \label{eq:SNR_SAR}
  SNR = \frac{P_t\,G_t\,G_r\,\lambda^3\,\sigma_0\,c}{128\,\pi^3\,R^3\,k\,T\,B\,F_n\,L_s\,v\,\sin\theta},
\end{equation}

\noindent{}which exhibits the distinct $1/R^3$ SAR factor.

From this equation it is clear that in order to improve SAR
performance, one should either increase the signal or decrease the
noise and losses (or both). The former asks for more power, more gain,
and greater wavelengths. The latter can be achieved, in principle, by
avoiding losses and increasing the velocity, $v$, which in turn requires
shorter pulses and an upgraded discrimination of the incoming signal.


\end{document}